\documentclass[sigconf, nonacm]{acmart}

\newcommand\vldbdoi{XX.XX/XXX.XX}
\newcommand\vldbpages{XXX-XXX}
\newcommand\vldbvolume{14}
\newcommand\vldbissue{1}
\newcommand\vldbyear{2020}
\newcommand\vldbauthors{\authors}
\newcommand\vldbtitle{\shorttitle} 
\newcommand\vldbavailabilityurl{URL_TO_YOUR_ARTIFACTS}
\newcommand\vldbpagestyle{plain}

\usepackage{listings}
\usepackage{xcolor}
\usepackage{graphicx}
\usepackage{subcaption}
\usepackage[ruled, vlined, linesnumbered]{algorithm2e}
\usepackage{amsthm}
\theoremstyle{definition}
\newtheorem{definition}{Definition}[section]
\usepackage{tablefootnote}

\definecolor{codegreen}{rgb}{0,0.6,0}
\definecolor{codegray}{rgb}{0.5,0.5,0.5}
\definecolor{codepurple}{rgb}{0.58,0,0.82}
\definecolor{backcolour}{rgb}{0.95,0.95,0.92}

\lstdefinestyle{mystyle}{
    backgroundcolor=\color{backcolour},   
    commentstyle=\color{codegreen},
    keywordstyle=\color{magenta},
    numberstyle=\tiny\color{codegray},
    stringstyle=\color{codepurple},
    basicstyle=\ttfamily\footnotesize,
    breakatwhitespace=false,         
    breaklines=true,                 
    captionpos=b,                    
    keepspaces=true,                 
    numbers=left,                    
    numbersep=5pt,                  
    showspaces=false,                
    showstringspaces=false,
    showtabs=false,                  
    tabsize=2
}

\usepackage{tikz}
\newcommand*\circled[1]{\tikz[baseline=(char.base)]{
            \node[shape=circle,draw,inner sep=0.8pt] (char) {#1};}}

\newcommand{\eat}[1]{}

\newcommand{\system}{{\fontfamily{lmtt}\selectfont RelServe}}
\newcommand{\relquery}{{\fontfamily{lmtt}\selectfont relQuery}}
\newcommand{\relqueries}{{\fontfamily{lmtt}\selectfont relQueries}}
\newcommand{\ProblemAbbr}{{Head-of-Line blocking}}

\begin{document}
\title{{\system}: Fast LLM Inference Serving on Relational Data}


\author{Xin Zhang}
\affiliation{
  \institution{HKUST}
}
\email{xzhanggb@ust.hk}

\author{Shihong Gao}
\affiliation{
  \institution{HKUST}
}
\email{sgaoar@connect.ust.hk}

\author{Yanyan Shen}
\affiliation{
  \institution{Shanghai Jiao Tong University}
}
\email{shenyy@sjtu.edu.cn}

\author{Haoyang Li}
\affiliation{
   \institution{PolyU}
}
\email{haoyang-comp.li@polyu.edu.hk}

\author{Lei Chen}
\affiliation{
  \institution{HKUST (GZ) and HKUST}
}
\email{leichen@cse.ust.hk}

\begin{abstract}
The use of Large Language Models (LLMs) for querying relational data has given rise to {\relquery}, a workload pattern that applies templated LLM calls to structured tables.
As {\relquery} services become more widely adopted in applications such as AI-powered spreadsheets, fast response times under concurrent query loads are increasingly important.
Unfortunately, current LLM engines face severe latency bottlenecks from Head-of-Line (HoL) blocking across three comparable inference phases: waiting, core running, and tail running. Existing static priority scheduling methods only address HoL blocking during the waiting phase, leaving two critical problems unsolved.
First, the absence of a priority update mechanism causes inaccurate prioritization and continued HoL blocking during core execution.
Second, suboptimal prefill-decode batching exacerbates HoL blocking in tail execution and worsens latency trade-offs between running and waiting {\relqueries}.
To address these problems, we propose {\system}, an optimized LLM engine for low-latency {\relquery} serving. {\system} features two core innovations: a Dynamic Priority Updater that continuously adjusts priorities while minimizing overhead via statistical approximations, and an Adaptive Batch Arranger that quantitatively evaluates candidate prefill and decode batches to minimize projected average latency.
Extensive experiments on four real-world datasets using LLMs ranging from 13B to 70B parameters show that {\system} reduces average serving latency by up to 3.1× compared to vLLM.
\end{abstract}

\maketitle

\pagestyle{\vldbpagestyle}
\begingroup\small\noindent\raggedright\textbf{PVLDB Reference Format:}\\
\vldbauthors. \vldbtitle. PVLDB, \vldbvolume(\vldbissue): \vldbpages, \vldbyear.\\
\href{https://doi.org/\vldbdoi}{doi:\vldbdoi}
\endgroup
\begingroup
\renewcommand\thefootnote{}\footnote{\noindent
This work is licensed under the Creative Commons BY-NC-ND 4.0 International License. Visit \url{https://creativecommons.org/licenses/by-nc-nd/4.0/} to view a copy of this license. For any use beyond those covered by this license, obtain permission by emailing \href{mailto:info@vldb.org}{info@vldb.org}. Copyright is held by the owner/author(s). Publication rights licensed to the VLDB Endowment. \\
\raggedright Proceedings of the VLDB Endowment, Vol. \vldbvolume, No. \vldbissue\ %
ISSN 2150-8097. \\
\href{https://doi.org/\vldbdoi}{doi:\vldbdoi} \\
}\addtocounter{footnote}{-1}\endgroup

\ifdefempty{\vldbavailabilityurl}{}{
\vspace{.3cm}
\begingroup\small\noindent\raggedright\textbf{PVLDB Artifact Availability:}\\
The source code, data, and/or other artifacts have been made available at \url{\vldbavailabilityurl}.
\endgroup
}

\section{Introduction}
\label{sec:intro}
\begin{figure}[t]
    \centering
    \includegraphics[width=\linewidth]{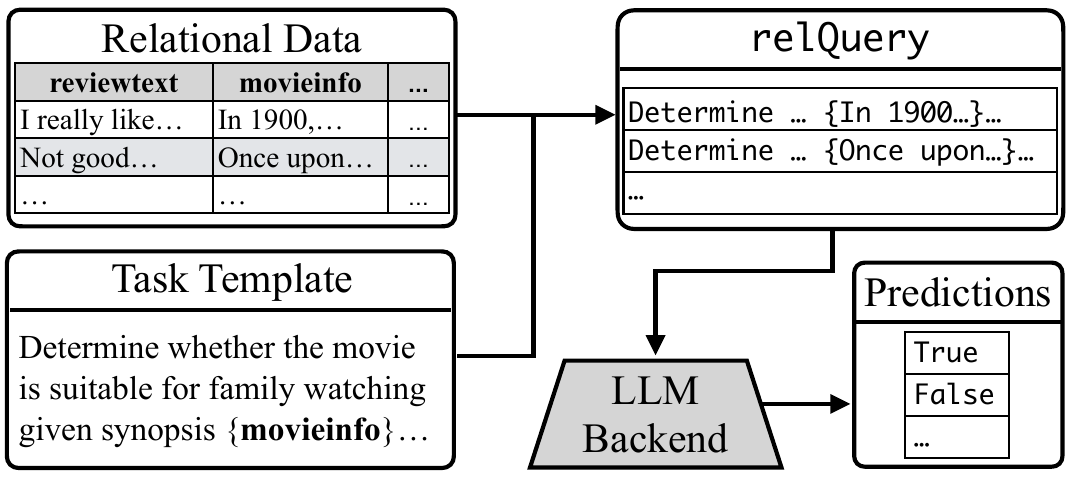}
    \vspace{-6ex}
    \caption{Example of {\relquery} service.}
    \label{fig:relquery}
    \vspace{-1ex}
\end{figure}

Recent developments across industry and academia demonstrate growing interest in applying Large Language Models (LLMs) to relational data. Many emerging applications require real-time semantic reasoning over structured data. For instance, AI-powered spreadsheets~\cite{sheetgpt,cellm,coefficient} now offer LLM-based cell functions capable of autofilling multiple rows~\cite{gptx_functions}. Major database vendors~\cite{databricks,redshift,bigquery} have integrated LLMs as built-in AI functions within SQL queries for streaming analytics~\cite{streaming_analytics}. Online business intelligence (BI) dashboards also leverage LLMs to generate insights and summaries over different data groups~\cite{TableauPulse}. 
These applications have given rise to a new class of workloads referred to as \textbf{\relquery}. As depicted in Figure~\ref{fig:relquery}, when users apply an LLM prediction function over an $N$-row table, they create a {\relquery} $R$ containing $N$ individual requests. Each request is formed by filling the task template with data from its corresponding row. The {\relquery} with $N$ requests is then sent to an LLM backend for processing.

Typically, a {\relquery} $R$ undergoes an \emph{iterative} inference procedure where each iteration executes either a prefill batch or a decode batch. A prefill batch processes a subset of requests from $R$ by computing intermediate representations for all input tokens and producing the first output token per request. A decode batch generates a subsequent output token for each request by leveraging the intermediate representations of input tokens and previously generated tokens.  
The {\relquery} $R$ completes when all its constituent requests finish processing, i.e., reaching a termination token or hitting the specified length limit per request. 

\emph{Low-latency} responses are critical for {\relquery} services, as users expect fast responses, such as when applying an LLM cell function to multiple spreadsheet rows. Delays in {\relqueries} degrade user experience and reduce satisfaction~\cite{li2015supporting,arapakis2014impact}. However, achieving low latency is challenging due to the prevalence of \emph{Head-of-Line (HoL) blocking}, a situation where short-running {\relqueries} are delayed behind earlier, long-running ones, significantly increasing overall response times. For example, the state-of-the-art LLM inference engine vLLM~\cite{vllm} incurs a prolonged average latency of \textbf{35 seconds} in a typical {\relquery} workload\footnote{Details in Section~\ref{sec:exp}. Setting: OPT-13B model, Rotten dataset, 1.0 {\relquery}/s.}. 
We later demonstrate how mitigating HoL blocking reduces this latency to \textbf{13 seconds}.

\begin{figure}[t]
    \centering
    \includegraphics[width=\linewidth]{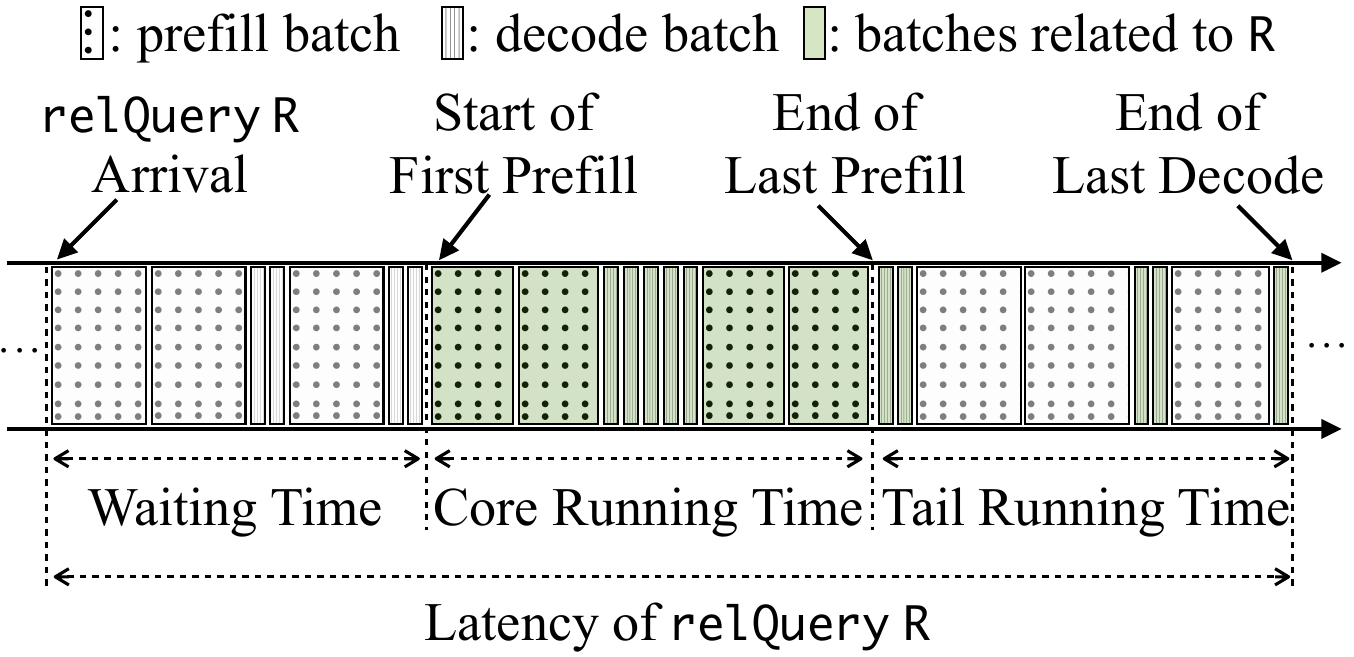}
    \vspace{-6ex}
    \caption{Trace profiling of vLLM and the three latency periods when serving a {\relquery} (X-axis: time).}
    \label{fig:three_latency_components}
    \vspace{-1ex}
\end{figure}

To systematically understand how HoL blocking affects the inference latency of a {\relquery} $R$, we decompose its inference process into three distinct periods as shown in Figure~\ref{fig:three_latency_components}: (1) \textbf{waiting time} spanning from its arrival to the initiation of the first prefill batch; (2) \textbf{core running time} covering the interval from the start of the first prefill batch to the completion of the final prefill batch; and (3) \textbf{tail running time} extending from the final prefill batch completion to the end of the last decode batch. 
Existing LLM serving systems primarily address HoL blocking during waiting time.
Specifically, LLM engines such as vLLM and Orca~\cite{orca, vllm, agrawal2024taming} adopt \textit{first-come-first-serve} (FCFS) scheduling, which processes requests strictly in their arrival order. Under FCFS, short-running {\relqueries} that arrive shortly after long-running ones are inevitably delayed, creating unnecessary queuing delays during the waiting time.
To address this issue, static priority scheduling strategies have been proposed~\cite{zheng2023response, aiops2024qiu, zhao2024alise, fu2024efficient}, which estimate the execution duration of each {\relquery} as a fixed priority value.
Requests are then scheduled in ascending priority value order\footnote{Note that a lower priority value indicates a higher actual priority in scheduling.}, allowing short-running {\relqueries} to bypass longer ones for earlier processing and thereby avoiding waiting-time HoL blocking.

However, our analysis reveals that \textbf{HoL blocking persists beyond waiting time}. With the increase of serving workload, HoL blocking accounts for 37\% to 63\% of latency during the core and tail running periods\footnote{See details about latency breakdown in Section~\ref{sec:exp:ablation}.}, which have comparable execution durations to the waiting time, due to different reasons.

First, the absence of a priority update mechanism accounts for HoL blocking during the core running time. Existing methods treat each {\relquery} as a fixed-priority task, ignoring both execution progress and context reuse opportunities. For example, a nearly-completed {\relquery} retains its original high priority value and can be preempted by newly arrived longer-running {\relqueries}, causing unnecessary delays. Moreover, existing methods fail to account for context reuse opportunities in {\relqueries}. 
Intermediate results from common templates and similar table values can be reused through prefix caching, meaning that {\relqueries} with significant prefix overlap execute more efficiently.
Incorporating prefix reuse into priority estimation would enable more accurate scheduling (see details in  Section~\ref{sec:bg:llm_infernce} and Section~\ref{sec:motivation:priority}). 

Second, suboptimal batch arrangement exacerbates HoL blocking during the tail running period. Existing methods, which are usually built atop modern inference engines like vLLM, prioritize prefill batches over decode batches to maximize parallelism and GPU utilization~\cite{vllm, agrawal2024taming}. 
However, this prefill-first strategy repeatedly interrupts decode batches, delaying completion of running {\relqueries} and extending tail running time.
We observe that the tail running period consumes comparable execution time to the core running period, despite processing far fewer tokens (often less than 10\% of total tokens). A na\"ive decode-first policy, while reducing the latency during tail execution, usually harms overall performance due to reduced parallelism and GPU utilization (see Section~\ref{sec:motivation:phase} for details).

In this paper, we propose \textbf{\system}, an efficient inference framework to address the HoL blocking problem in {\relquery} services. {\system} dynamically updates the priority value of each {\relquery} in every iteration, considering both its execution progress and reusable intermediates in the engine. This ensures that priority values remain accurate and responsive to the changing runtime conditions.
Furthermore, {\system} adaptively arranges the execution order of prefill and decode batches, rather than rigidly prioritizing one over the other, to mitigate HoL blocking in the tail running time and minimize the average latency.
However, implementing {\system} requires addressing the following two key challenges.

\noindent \textbf{Challenge 1: High Overhead of Priority Updates.} Accurate prioritization requires real-time awareness of both the execution state and the number of reusable intermediates. However, obtaining this information is non-trivial and time-consuming. For example, computing the exact number of reusable intermediates may require checking thousands of requests against a prefix cache containing tens of thousands of intermediates. Because the execution state can change every few milliseconds, performing such an expensive matching operation at each iteration introduces substantial overhead, which can negate the performance gains from prioritization.

\noindent \textbf{Challenge 2: Complex Latency Trade-offs in Batch Scheduling.} The decision on the execution order of prefill and decode batches involves complex latency trade-offs between the currently executing {\relqueries} and the waiting ones. For example, while prioritizing the prefill batch from a waiting {\relquery} reduces its waiting time with improved parallelism and GPU utilization, the tail running time of the currently executing {\relquery} is inevitably extended (see details in Section~\ref{sec:motivation:phase}). Balancing these conflicting objectives is challenging because the optimal schedule continually shifts with the evolving states of running and waiting {\relqueries}.

To tackle these challenges, {\system} consists of two key components: the Dynamic Priority Updater (DPU) and the Adaptive Batch Arranger (ABA). First, DPU monitors the completion information and the prefix cache content and dynamically updates the priority value of all {\relqueries} in each iteration to ensure an accurate priority estimation. To reduce the overhead, DPU adopts several approximation techniques such as historical priority reusing and prefix cache sampling based on statistics. 
Second, ABA addresses the complex latency trade-offs by holistically and quantitatively evaluating all {\relqueries} in the system.
In each iteration, ABA constructs both a candidate prefill batch and a candidate decode batch, and then computes the projected latency change for every {\relquery} in the inference engine under each candidate. ABA finally selects the candidate that yields the lowest average latency.

We implement {\system} on top of the state-of-the-art LLM serving system vLLM~\cite{vllm} to take advantage of existing advances in the LLM ecosystem. Our implementation extends vLLM with support for dynamic priority updating and adaptive prefill/decode batch scheduling mechanisms.
We evaluate the performance of {\system} on three LLM models with 13B, 32B, and 70B parameters, four real-world datasets, and five representative types of {\relqueries} across various serving settings. Experimental results demonstrate that {\system} reduces average latency by up to 3.1 and 1.6 times compared to vLLM and state-of-the-art static priority scheduling methods. Below summarizes our major contributions.

\begin{itemize}
    \item We formally define {\relquery}, a new LLM serving workload on relational data, and conduct a comprehensive analysis that identifies three distinct latency periods in {\relquery} serving where HoL blocking occurs.
    \item We propose {\system}, a novel LLM inference framework that addresses HoL blocking in {\relquery} services. It incorporates two key components: a Dynamic Priority Updater that provides accurate priority estimation by considering execution progress and content reuse opportunity, and an Adaptive Batch Arranger that coordinates prefill and decode batches to minimize average latency.
    \item We conduct extensive experiments to validate the efficiency of {\system} on real-world datasets, diverse workload patterns, and representative LLM models ranging from 13B to 70B parameters. 
    Experimental results demonstrate that {\system} reduces average latency by up to 3.1× compared to vLLM and up to 1.6x compared to state-of-the-art static priority scheduling methods.
\end{itemize}

\section{Background}  
\label{sec:bg}

In this section, we formally define {\relquery}, describe its serving properties, and introduce fundamental concepts of LLM inference. Table~\ref{tb:notation} summarizes the key notations used throughout the paper.

\begin{table}[t]
    \centering
    \caption{Key notations and their descriptions.}
    \label{tb:notation}
    \vspace{-3ex}
    \begin{tabular}{c|c}
    \toprule
    Notation & Description \\
    \midrule
    \midrule
    $R$ & a {\relquery}  \\
    \midrule
    $r$ & a request  \\
    \midrule
    $p$ & a prefill batch  \\
    \midrule
    $d$ & a decode batch  \\
    \midrule
    $Q^+_t$ & the running queue at iteration $t$  \\
    \midrule
    $Q^-_t$ & the waiting queue at iteration $t$  \\
    \midrule
    $\mathcal{R}_t^+$ & all {\relqueries} in the running queue at iteration $t$  \\
    \midrule
    $\mathcal{R}_t^-$ & all {\relqueries} in the waiting queue at iteration $t$ \\
    \midrule
    $R_t$ & $R$ excluding completed requests at iteration $t$  \\
    \midrule
    $\alpha^p/\alpha^d$ & slope of token count vs. batch duration linearity  \\
    \midrule
    $\beta^p/\beta^d$ & intercept of token count vs. batch duration linearity  \\
    \midrule
    $\mathtt{Prio}(\cdot)$ & the priority value of a request/{\relquery} \\
    \midrule
    $tok(\cdot)$ & number of input tokens in a request/batch \\
    \midrule
    $utok(\cdot)$ & num. of uncached input tokens in a request/batch \\
    \midrule
    $req(\cdot)$ & number of requests in a batch/{\relquery} \\
    \midrule
    $OL(\cdot)$ & output length of the request in a {\relquery}/batch  \\
    \bottomrule
    \end{tabular}  
\end{table}

\subsection{Definition and Serving of {\relquery}}
\label{sec:bg:relquery}

We focus on the effective utilization of LLMs for processing relational data through {\relqueries}. These queries constitute a category of workloads where LLMs perform templated tasks over rows of relational datasets, which can be formally defined as follows.

\begin{definition}[{\relquery} $R$]
Consider a relational table $T=\{C,S\}$ with schema $C=\{c_1, c_2, \ldots, c_m\}$, where each $c_j$ is an attribute, and $S = \{s_1, s_2, \ldots, s_n\}$ denotes the collection of rows. A \textit{task template} $\zeta$ is a text string containing attribute names $\{\phi_1, \phi_2, \ldots, \phi_k\}$, where $\phi_j \in C, j \in [1,k], k \le m$. We define {\relquery} $R$ as:
\begin{equation}
    R = \text{\fontfamily{lmtt}\selectfont relQuery}(T, \zeta) = \{r_i | i\in[1,n]\} =  \{\zeta[s_i] \space | \space s_i\in S\}.
\end{equation}
Here $r_i=\zeta[s_i]$ denotes a request in $R$ instantiated from template $\zeta$ through the substitution of each $\phi_j$ with the corresponding attribute values $s_i[\phi_j]$ from row $s_i$, i.e., $\zeta[\phi_1 \mapsto s_i[\phi_1], \dots, \phi_k \mapsto s_i[\phi_k]].$

\end{definition}

We illustrate the {\relquery} construction process with the rating prediction task on the Rotten Tomatoes table~\cite{rotten_ds}, which contains two attributes: \verb|{reviewcontent}| storing user reviews and \verb|{movieinfo}| holding movie synopses. The task template is \textit{"Predict rating (1 to 5) for the movie \{movieinfo\} given the user's review \{reviewcontent\}. Output only digit and nothing else"}. For each row, we generate an LLM request by filling the corresponding values of \verb|{reviewcontent}| and \verb|{movieinfo}| into the template. All such requests collectively constitute the {\relquery} for this task.

When submitting a {\relquery} $R$, users can specify a maximum output length $OL(R)$. For instance, in rating prediction tasks, each LLM request should produce a single digit output. Consequently, setting the maximum output length to 5 tokens is adequate when accounting for additional special tokens like begin-of-sentence or end-of-sentence markers. {\relqueries} for other tasks such as summarization or open-ended question answering typically require longer output lengths, often ranging from 50 to 100 tokens.

\vspace{-1ex}
\begin{definition}[{\relquery} Inference Latency]

Given a {\relquery} $R$ whose inference involves some prefill batches $\mathcal P=\{p_1, \ldots, p_m\}$ and decode batches $\mathcal D=\{d_1, \ldots, d_n\}$. Let $arrival(R)$ denote the arrival timestamp of $R$. For any batch $b\in\mathcal P\cup\mathcal D$, we use $start\_ts(b)$ and $end\_ts(b)$ to denote its start and end timestamps, respectively.
The total latency of $R$ can be divided into three parts: the waiting time during which $R$ remains in the waiting queue, the core running time when all prefill batches of $R$ and a portion of $R$'s decode batches are executed, and the tail running time when the remaining decode batches of $R$ are executed, i.e.,
\begin{equation}
\begin{aligned}
    \text{Waiting\_Time}(R) &= start\_ts(p_1) - arrival(R), \\
    \text{Core\_Running\_Time}(R) &= end\_ts(p_m) - start\_ts(p_1), \\
    \text{Tail\_Running\_Time}(R) &= end\_ts(d_n) - end\_ts(p_m).
\end{aligned}
\end{equation}

\end{definition}

\noindent \textbf{High Cost of {\relquery} Serving.} While integrating LLMs with relational data brings semantic reasoning capabilities far beyond those of basic data operations, LLMs' inference can be orders of magnitude slower than traditional spreadsheet or relational operations on standard hardware. For example, Microsoft Excel and DuckDB easily process MB-level and GB-level of data per second on laptops~\cite{excel,duckdb}, while serving a 13B OPT model on a server equipped with an A100 GPU processes data at approximately 30KB per second.
Given this substantial performance disparity, enabling fast {\relquery} processing is critical. 

\noindent \textbf{Online versus Offline {\relquery} Serving.} {\relqueries} are widely applicable to both online and offline settings, where optimization targets differ. First, \textit{low latency} is critical for online applications like interactive data exploration, streaming analytics, and BI dashboards, where users expect instant responses~\cite{gptx_functions,streaming_analytics,TableauPulse,li2015supporting,arapakis2014impact,zou2013flexquery, agarwal2012blink, chaudhuri2011overview}. Minimizing latency in {\relquery} services is essential since delays degrade user experiences. Second, \textit{high throughput} is desired for offline applications like log processing and offline ranking over historical data, where users can tolerate hours or even days of latency. These tasks are typically performed in batch at a low frequency, for example, as routine tasks performed daily or weekly. Consequently, it is more beneficial to optimize the throughput. We summarize these distinctions in Table~\ref{tb:comparison}. 
This paper focuses on optimizing latency for online {\relquery} serving, motivated by both the prevalence of online applications and their time-sensitive nature.

\subsection{LLM Inference}
\label{sec:bg:llm_infernce}
Mainstream LLMs such as GPT~\cite{achiam2023gpt}, Llama~\cite{touvron2023llama}, and Qwen~\cite{bai2023qwen} are built using decoder-only Transformer~\cite{vaswani2017attention} architectures. These models typically contain an Embedding Layer, a stack of Transformer Layers, and an Output Layer. Given an input token sequence $TS=(t_1, t_2, ..., t_n)$, the Embedding Layer maps input tokens to vectors $\mathbf X=(\mathbf x_1, \mathbf x_2, ..., \mathbf x_n)$. This vector sequence is then processed by Transformer Layers, which capture inter-token dependencies and produce hidden representations $\mathbf H=(\mathbf h_1, \mathbf h_2, \dots, \mathbf h_n)$. The Output Layer uses the final hidden representation $\mathbf h_n$ to compute logits for predicting the next token $t_{n+1}$. 
This autoregressive decoding continues until a special end-of-sentence (EOS) token is produced or a maximum length is reached.
At the core of each Transformer layer lies the self-attention module as detailed below.

\noindent \textbf{Self-Attention Module}. In the $\ell$-th Transformer Layer, the self-attention module first transforms its input vectors $\mathbf X=(\mathbf x^\ell_1, \mathbf x^\ell_2, ..., \mathbf x^\ell_n)$ into the \textit{query}, \textit{key}, and \textit{value} vectors: 
\begin{equation}
    \label{eq:qkv}
    \mathbf q^\ell_i = \mathbf W^\ell_q \mathbf x^\ell_i, \quad 
    \mathbf k^\ell_i = \mathbf W^\ell_k \mathbf x^\ell_i,
    \quad
    \mathbf v^\ell_i = \mathbf W^\ell_v \mathbf x^\ell_i.
\end{equation}

Next, it computes attention scores $\alpha^\ell_{ij}$for each \textit{query} vector $\mathbf q^\ell_i$ by measuring its similarity with all previous \textit{key} vectors:
\begin{equation}
    \label{eq:attention_score}
    \alpha^\ell_{ij} = \frac{\exp \left({\mathbf q^{\ell}_{i}}^{\top}\mathbf k^\ell_j / \sqrt{d} \right)}{\sum_{j=1}^i \exp \left( {\mathbf q^{\ell}_{i}}^{\top}\mathbf k^\ell_j / \sqrt{d} \right) }.
\end{equation}

Finally, the output vector $\mathbf o^\ell_i$, which is used to calculate $\mathbf h_i$ later, is a weighted sum of the \textit{value} vectors $\{\mathbf v^\ell_j\}_{j=1}^i$ w.r.t. attention scores:

\begin{equation}
    \label{eq:attention_out}
    \mathbf o^\ell_i = \mathbf W^\ell_o \sum_{j=1}^i \alpha^\ell_{ij} \mathbf v^\ell_{j}.
\end{equation}

\begin{table}[t]
    \caption{Comparison of online and offline {\relquery} usage.}
    \vspace{-3ex}
    \centering
    \begin{tabular}{rll}
    \toprule
     & Online & Offline \\
    \midrule
    \midrule
    Urgency & High & Low \\
    Usage Pattern & Interactive  & Batch Processing \\
    Typical Latency & Seconds & Hours to Days \\
    Optimization Target & Low Latency & High Throughput \\
    \bottomrule
    \end{tabular}
    \label{tb:comparison}
\end{table}


\noindent \textbf{KV Cache}. The computation outlined in Equations~(\ref{eq:qkv})-(\ref{eq:attention_out}) reveals that the \textit{key} vectors $\{\mathbf k^\ell\}$ and the \textit{value} vectors $\{\mathbf v^\ell\}$ at each Transformer Layer of the preceding tokens are fixed but repeatedly calculated when predicting future tokens. To avoid this redundant computation, existing LLM serving frameworks~\cite{vllm, agrawal2024taming} cache previously computed \textit{key} and \textit{value} vectors, referred to as the KV cache, thereby accelerating the inference of one request.

\noindent \textbf{Prefill and Decode Phases}. 
LLM inference for a single request proceeds in two phases: prefill and decode. In the prefill phase, the model processes all input tokens to build the initial KV cache and generates the first output token. In the decode phase, the model leverages the existing KV cache to sequentially generate additional output tokens, one at a time, until an end-of-sequence (EOS) token is produced or a length constraint is reached. Consequently, each request undergoes one prefill phase followed by multiple decode phases (one per generated token).
Modern LLM serving systems batch the prefill/decode phases across multiple requests to improve throughput. However, these two phases exhibit different computational characteristics~\cite{vllm, agrawal2024taming}. The decode phase is memory-bound: it performs relatively little computation per memory access, leading to GPU underutilization when batch sizes are small. In contrast, the prefill phase is compute-bound: it performs substantial computation, resulting in high GPU utilization regardless of batch size.

\noindent \textbf{Prefix Cache}. When multiple requests share a common prefix (e.g., task instructions), inference can be further accelerated by reusing the KV cache from earlier requests. Current LLM serving systems, such as vLLM, maintain a global prefix cache that stores previously computed KV cache entries across requests. Upon receiving a new request, the system identifies the matching prefix in the cache and skips the corresponding computation. Prefix cache is maintained with an LRU eviction policy to maximize utility under memory constraints. The prefix cache is updated online as new requests are processed. A high prefix cache hit ratio substantially reduces inference time by avoiding redundant computation for identical prefix tokens. However, the overall benefit depends on the degree of prefix overlap in the workload.

\section{Motivation}  
\label{sec:motivation}

This section presents experimental findings\footnote{Experiment setting: Amazon dataset, OPT-13B model, 1.0 {\relquery} coming per second, randomly sampled 100 {\relqueries}.} that reveal different underlying reasons for HoL blocking in the core and tail running periods. We first pinpoint two dynamic factors ignored by static priority methods, causing HoL blocking in the core running period. Then we discuss the prolonged tail running time caused by the na\"ive prefill and decode arrangement in existing serving systems~\cite{vllm,fastertransformer}.

\subsection{Influence of Execution Progress and Context Reuse on Priority} 
\label{sec:motivation:priority}

\begin{figure}[t]
    \centering
    \includegraphics[width=0.9\linewidth]{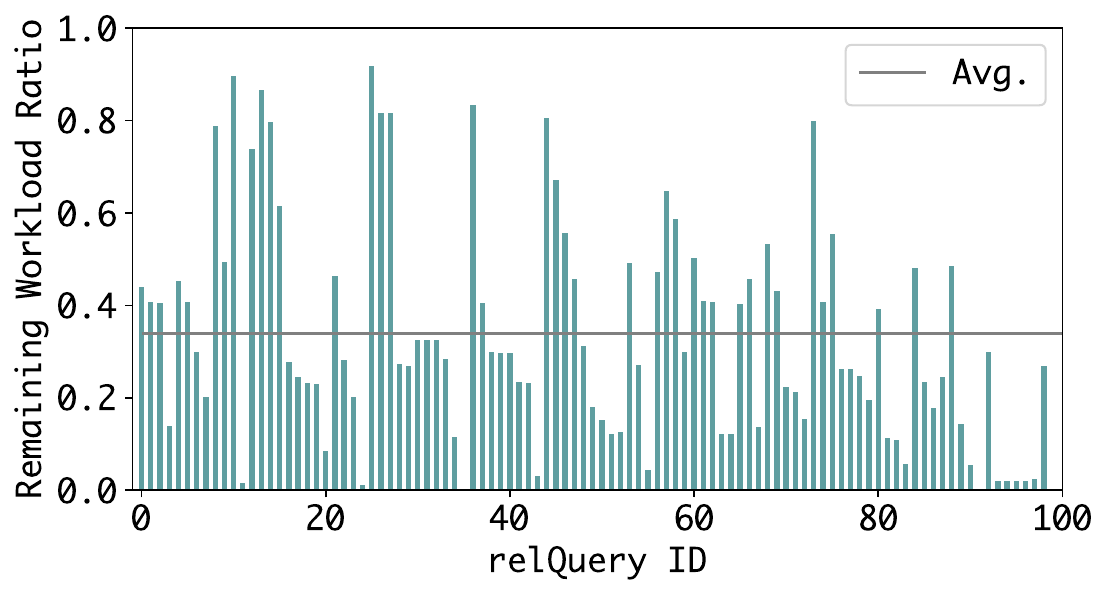}
    \vspace{-4ex}
    \caption{Remaining workload ratio of each running {\relquery} when the next {\relquery} arrives.}
    \label{fig:staleness}
\end{figure}

To mitigate {\ProblemAbbr}, static priority scheduling methods~\cite{zheng2023response, aiops2024qiu, zhao2024alise, fu2024efficient} arrange requests according to their estimated execution times. And the priority value of an individual request $r$, denoted as ${\texttt{ReqPrio}(r)}$, is estimated solely based on its input and output token counts using predefined functions $L^1$ and $L^2$:
\begin{equation}
\label{eq:request_priority}
    \texttt{ReqPrio}(r) = L^{1} (tok(r)) + L^{2}(OL(r)),
\end{equation}
where $tok(r)$ represents the number of input tokens and $OL(r)$ denotes the number of output tokens. The choice of $L^1$ and $L^2$ varies across methods, using constant or identical mappings~\cite{fu2024efficient, aiops2024qiu, zheng2023response}, or simple linear functions~\cite{zhao2024alise}.

In the context of {\relquery} serving, the priority value $\texttt{Prio}(R)$ for a {\relquery} $R$ is calculated by summing $\texttt{ReqPrio}(r_i)$ of all constituent requests. This calculated $\texttt{Prio}(R)$ is then shared by all requests $r_i$ within $R$.
\begin{equation}
    \texttt{Prio}(R) = \texttt{Prio}(r_i) = \sum_{r_i \in R} \texttt{ReqPrio}(r_i).
\end{equation}

However, existing methods compute the priority values of requests once and keep them \textbf{fixed} throughout the inference process. 
This results in inaccurate estimations for {\relquery} workloads due to the ignorance of two key dynamic factors: the execution progress of {\relquery} and available context reuse opportunities.

\noindent \textbf{Execution Progress.} Traditional LLM requests are independent in serving, but requests in one {\relquery} are interdependent because they together contribute to one latency of the {\relquery}. If a {\relquery} is partially completed, the {\relquery} has less remaining workload and should have a smaller priority value. Therefore, when some requests in a {\relquery} are completed, other requests should be informed about the completion information, and the priority values of the remaining requests should be reduced accordingly. We demonstrate the importance of this priority update by visualizing the remaining workload ratio of the running {\relquery} when a new {\relquery} arrives, as shown in Figure~\ref{fig:staleness}. On average, the remaining workload accounts for only 34\% of the total workload, which means that the actual priority value of the {\relquery} should be reduced to 34\% of its original fixed priority value. Without the awareness of execution progress and the dynamic priority update, the priority value of the running {\relquery} is mistakenly magnified, leading to HoL blocking and prolonged latency.

\begin{figure}[t]
    \centering
    \includegraphics[width=0.9\linewidth]{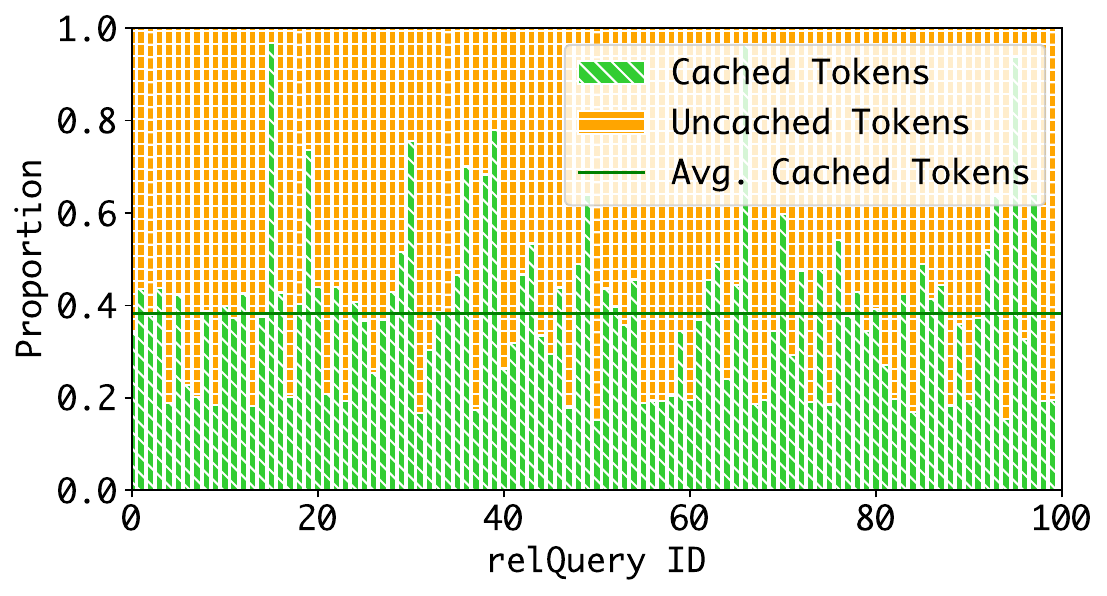}
    \vspace{-4ex}
    \caption{Cached \& uncached tokens in different {\relqueries}.}
    \label{fig:prefix_cache}
\end{figure}

\noindent \textbf{Shared Prefixes.} There are some shared contexts between requests within the same {\relquery}. For example, requests of a {\relquery} are instantiated from the same template, and the values of the same attribute are similar. As we have introduced in Section~\ref{sec:bg:llm_infernce}, the prefix cache stores the computed KV cache of previous requests, part of which can be reused to avoid redundant computation. Formally, the prefix cache reduces the number of input tokens $tok(r)$ that need to be computed in Equation~\ref{eq:request_priority}. We inspect the number of prefix cache miss tokens, which requires prefill computation, and the number of prefix cache hit tokens, which requires no computation, as shown in Figure~\ref{fig:prefix_cache}. The prefix cache hit ratios of different {\relqueries} are diverse with an average of 38\%. Therefore, ignoring shared prefixes causes varied levels of overestimation in {\relquery}'s priority, leading to HoL blocking in scheduling.

\vspace{-1ex}
\subsection{Prefill and Decode Arrangement}
\label{sec:motivation:phase}

\begin{figure}[]
    \centering
    \includegraphics[width=0.9\linewidth]{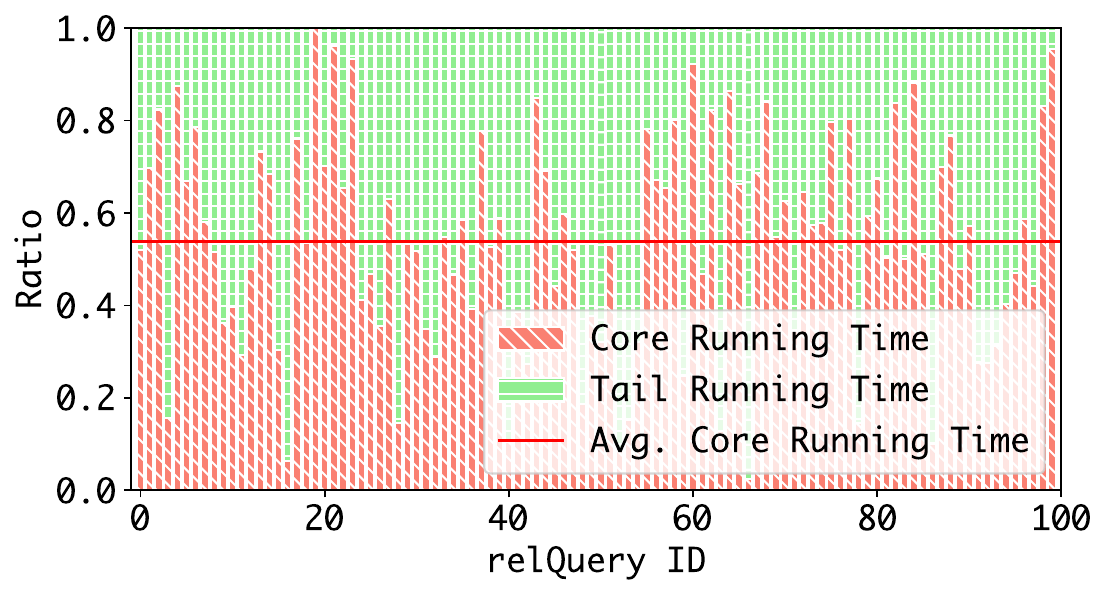}
    \vspace{-4ex}
    \caption{Normalised core\&tail running times of {\relqueries}.}
    \label{fig:inter_intra_breakdown}
\end{figure}

As we have discussed in Section~\ref{sec:intro}, vLLM na\"ively prioritizes the prefill phase over the decode phase, leading to a long tail running time in {\relquery} serving. To demonstrate this problem, we profile the vLLM inference and compare the core running time and the tail running time of each {\relquery} in Figure~\ref{fig:inter_intra_breakdown}.

We observe that the core and tail running times are close (54:46), although only 8\% of the tokens are processed during the tail running time, revealing severe HoL blocking that extends the tail running time. This problem is caused by the strict prefill prioritization strategy of vLLM, which postpones the completion of the last few decode batches of the running {\relquery} and prioritizes the prefill batches of the next {\relquery} for parallel decoding.

\begin{figure*}[h!]
    \centering
    \includegraphics[width=\textwidth]{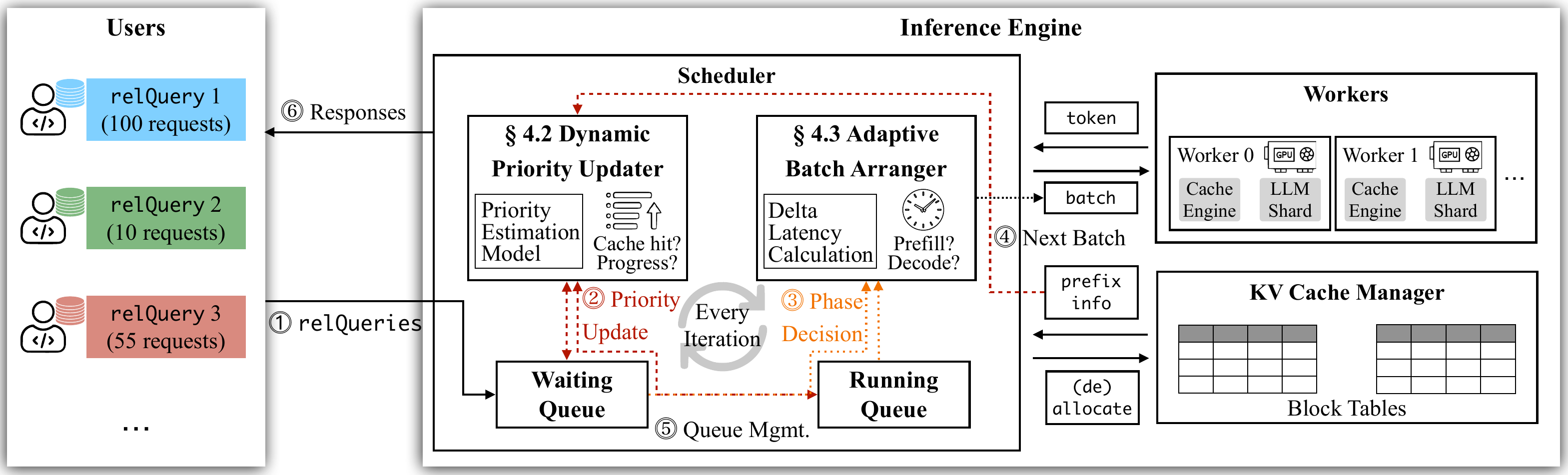}
    \vspace{-6ex}
    \caption{Overview of {\system}.}
    \label{fig:arch}
    \vspace{-2ex}
\end{figure*}

This prefill prioritization arrangement extends the latency of the current {\relquery} with later completion, but reduces the latency of the next {\relquery} for parallel decoding. A straightforward fix to the prolonged tail running time problem is the decode prioritization arrangement, which finishes the decode batches of the running {\relquery} before executing later prefill batches, thus minimizing the tail running time of the running {\relquery}. However, decode prioritization could increase the latency of next {\relqueries} due to the lack of combined decoding. In Section~\ref{sec:exp}, we will show that both prefill prioritization and decode prioritization are suboptimal.

In fact, the relative benefits of (1) earlier completion for running {\relqueries} and (2) increased decoding parallelism for later {\relqueries} depend on the prefill and decode batches involved and all {\relqueries} in the inference engine. Existing na\"ive prefill and decode arrangements cannot optimize this latency trade-off between prefilled {\relqueries} and yet-to-prefill {\relqueries}. This inspires us to develop an adaptive prefill and decode arrangement.

\section{Fast LLM Inference Serving for RelQueries}
\label{sec:method}
\subsection{\system~Overview}

Figure~\ref{fig:arch} depicts the overview of our {\system} framework for serving {\relqueries} through LLMs. At the core of {\system} lies the \textbf{Scheduler}, which functions as the primary coordinator to optimize inference efficiency through the following procedure.

\circled{1} \textbf{Initial {\relquery} Processing.} The Scheduler continuously receives {\relqueries} from multiple users. When a {\relquery} arrives, its constituent requests are placed in the waiting queue. The Scheduler then initiates iterative LLM inference processing on these queued requests. Each iteration executes either a prefill batch or a decode batch of requests.

\circled{2} \textbf{Priority-based Request Ordering.} At the beginning of each iteration, the Scheduler invokes the Dynamic Priority Updater to compute \emph{priority values} for all requests in the waiting queue and running queue using a Priority Estimation Model. The requests are arranged in ascending order with respect to priority values, where lower priority values indicate higher urgency. The highest-priority requests are selected for the next prefill batch. 

\circled{3} \textbf{Adaptive Batch Configuration.} The Scheduler utilizes the Adaptive Batch Arranger to determine the optimal batch type for each iteration, i.e., a prefill or a decode batch. This decision balances latency trade-offs. Specifically, prioritizing prefill batches reduces latency for yet-to-prefill \relqueries, while prioritizing decode batches accelerates the completion of already prefilled \relqueries. A delta latency prediction model guides these scheduling decisions to minimize average latency.

\circled{4} \textbf{Batch Processing.} The Scheduler dispatches the configured batch along with the allocated KV cache space to GPU workers for execution. GPU workers perform LLM forward computation on the assigned batch, maintain KV cache for each transformer layer, and return predicted tokens to the Scheduler. 

\circled{5} \textbf{Queue State Management.} After each iteration completes, the Scheduler updates queue states based on the processing results. When a prefill batch is processed, all constituent requests are transferred from the waiting queue to the running queue to begin their decode phase later. When a decode batch is processed, requests that have completed, i.e., reaching the end-of-sequence token or the length limit, are removed from the running queue. The priority recalculation process operates solely on requests remaining in either the waiting or running queues, excluding completed requests.

\circled{6} \textbf{Response Delivery.} The Scheduler provides responses to users upon completion of all requests within a {\relquery}.

The subsequent sections elaborate on the details of two key components in \system, namely the Dynamic Priority Updater that aims to avoid HoL blocking during the core running time and the Adaptive Batch Arranger that aims to optimize the latency trade-off in the tail running time.

\subsection{Dynamic Priority Updater}
\label{sec:method:dpu}
The Dynamic Priority Updater (DPU) enables accurate and computationally efficient priority estimation for {\relqueries}. 
	The core component in the DPU is the Priority Estimation Model (PEM), which estimates the execution duration of each {\relquery} by simulating its processing workflow. The DPU uses these duration estimates to update the corresponding priority values. 
	To reduce computational overhead, the DPU selectively reuses historical priority values when request characteristics remain stable, eliminating redundant calculations while preserving estimation accuracy.
	Formally, at the beginning of iteration $t$ ($t\geq 1$), for each $R_t$ in the engine, namely an unfinished {\relquery} excluding requests that have completed before the $t$-th iteration, the DPU either computes a new priority value via PEM or reuses the previously calculated value $\mathtt{Prio}(R_{t-1})$. That is,
	\begin{equation}
		\texttt{Prio}(R_t) = 
		\begin{cases}
			\texttt{Prio}(R_{t-1}), &\text{ if } t>1 \wedge reuse(R_t, R_{t-1}), \\
			\text{PEM}(R_t), &\text{ otherwise.}
		\end{cases}
	\end{equation}
    
Next, we describe the necessity of iteration-level priority updates, the details of the Priority Estimation Model, and the reuse mechanism for computational efficiency.  

\noindent \textbf{Iteration-level Priority Update.} Each inference iteration, which either executes a prefill or a decode batch, changes the runtime condition and impacts the request priorities. Specifically, the prefill execution updates the prefix cache, which reduces the workload of the remaining requests in the same {\relquery} via KV cache reuse. The decode execution generates the next tokens for requests in the batch. 
To capture these evolving runtime conditions, we adopt an iteration-level priority update mechanism that recalculates and updates priority values of requests at each iteration.

\noindent \textbf{{\relquery} Priority Estimation.} The priority value of a {\relquery} is determined by its execution duration, as prioritizing shorter {\relqueries} helps reduce average latency~\cite{blazewicz1996job}. 
Each {\relquery} consists of multiple requests that are processed in batches during inference. We find that the execution time of each (prefill or decode) batch can be reliably predicted based on the number of tokens contained in the batch. Therefore, we use the batch as an intermediate unit for estimating the execution time of a {\relquery}.

\begin{figure}[t]
    \centering
    \includegraphics[width=\linewidth]{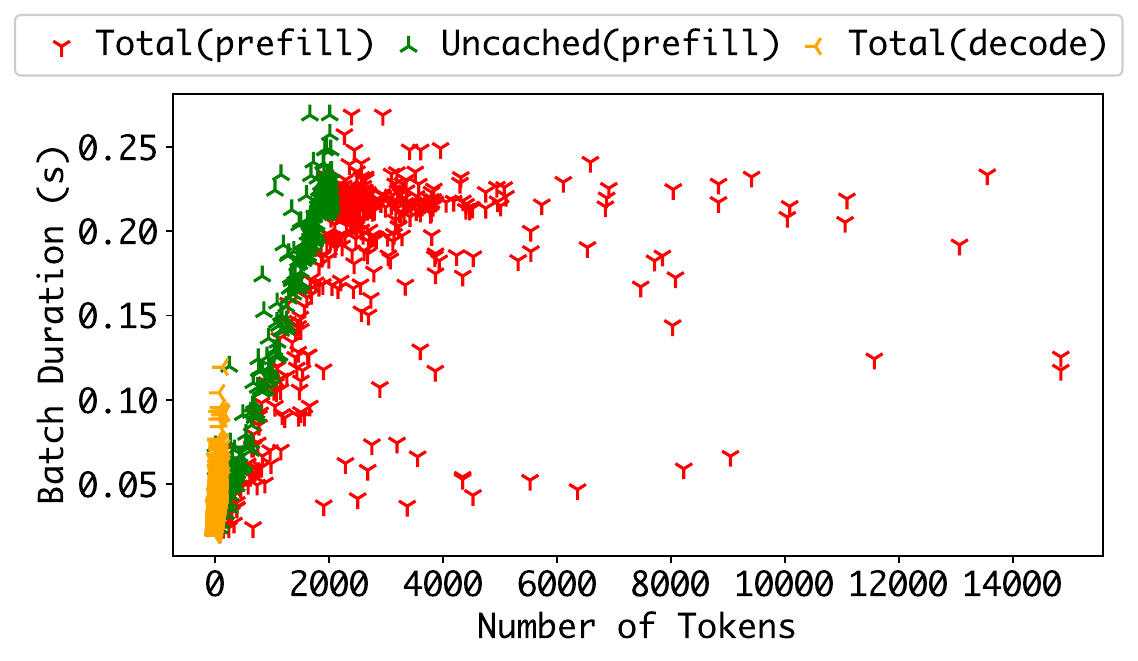}
    \vspace{-6ex}
    \caption{Relationship between the number of tokens in the batch and the prefill/decode batch duration.}
    \label{fig:linear_tokens_time}
\end{figure}

\noindent \textbf{Relationship Between the Number of Tokens and the Batch Duration.} To predict the duration of a batch, we use the number of tokens that determines the computation required by the batch. While previous works~\cite{zheng2023response,aiops2024qiu,zhao2024alise,fu2024efficient} predict the batch execution time with the total number of tokens in the batch, they ignore the dynamic prefix cache, leading to a weak linear relationship for the prefill batch. We use \textit{Total (prefill)} to represent all tokens in the prefill batch, \textit{Uncached (prefill)} to represent tokens that are not cached by the prefix cache, and \textit{Total (decode)} to represent all tokens in the decode batch. As shown in Figure~\ref{fig:linear_tokens_time}, we observe that by only accounting for uncached tokens that need to be practically computed, we obtain a much stronger linearity between the number of tokens and the prefill batch duration. Therefore, we use the following two linear functions $L^{prefill}$ and $L^{decode}$ to predict the duration of a prefill batch $p$ and a decode batch $d$:

\begin{equation}
\label{eq:batch_duration_function}
    \begin{aligned}
       L^{prefill}(p) &= \alpha^p\times utok(p)\ + \beta^p,\\
       L^{decode}(d) &= \alpha^d\times req(d) + \beta^d.
    \end{aligned}
\end{equation}

Recall that $utok(p)$ stands for the number of uncached tokens in the prefill batch $p$ and $req(d)$ means the number of requests in the decode batch $d$ as in Table~\ref{tb:notation}. The $\alpha^p/\alpha^d$ is the slope and the $\beta^p/\beta^d$ is the intercept, whose values are obtained by linearly fitting data points collected from several offline runs.

\noindent \textbf{Batch Decomposition of the {\relquery}.} Since we can predict the duration of each batch, we only need the batch composition of a {\relquery} to derive its execution duration. We present how to decompose the execution of a {\relquery} $R_t$ into a series of prefill batches $\mathcal P$ and decode batches $\mathcal D$ as in Algorithm~\ref{algo:batch_decompose}. 

The decomposition to prefill/decode batches must comply with the user-defined constraints, namely the maximal number of tokens in a prefill batch \verb|max_num_batched_tokens|, the maximal number of requests in a decode batch \verb|max_num_seqs|, and the maximal number of tokens on the GPU \verb|cap|. 

Given a request $r_i\in R_t$, we add it to the current prefill batch $p$ and decode batch $d$ if all constraints are met (Line 11). If adding $r_i$ violates the prefill batch size constraint, it means that $p$ is full; therefore, we add $p$ into $\mathcal P$ and clear $p$ for later prefill (Line 9-10). We continue prefilling until the violation of the constraint on the total number of tokens or the number of requests, when we execute and finish all requests on the GPU by adding $p$ to $\mathcal P$ and adding multiple $d$ to $\mathcal D$ (Line 4-8). Finally, the leftover batches are also added (Line 12-15). If a request $r_i$ in $R_t$ has already been prefilled and decoded for several iterations, we set $utok(r_i)$ to 0 and remove this $r_i$ from the corresponding prefill/decode batches.

\begin{algorithm}[t]
\caption{Batch Decomposition}
\label{algo:batch_decompose}
\DontPrintSemicolon
\KwIn{a {\relquery} containing $n$ requests $R_t=\{r_1, ...,r_n\}$, the number of computed tokens in KV Cache Manager constraint $cap$, the decode batch size constraint $\text{max\_num\_seqs}$ as $mns$, the prefill batch size constraint $\text{max\_num\_batched\_tokens}$ as $mnbt$.}
\KwOut{List of prefill batches $\mathcal P$ and decode batches $\mathcal D$.}
\SetKwBlock{Begin}{function}{end function}

\Begin($\text{batch\_decompose} {(} R_t {)}$) 
{
    $\mathcal P \gets \emptyset$, $\mathcal D \gets \emptyset$, $p \gets \emptyset$, $d \gets \emptyset$, $accum \gets 0$\;
    \For{$i = 1,...,n$}{
        \If{$utok(r_i) +accum > cap \text{ or } req(d)+1 >mns$}{
            $\mathcal P$.add($p$)\;
            \For{$o = 1,...,OL(R_t)$}{
                $\mathcal D$.append($d$) \tcp*{decode to completion}
            }
            $p \gets \emptyset$, $d \gets \emptyset$
        }
        \If{$utok(r_i) +utok(p) > mnbt$}{
            $\mathcal P$.add($p$), $p \gets \emptyset$\;
             
        }
        
        $p$.add($r_i$), $d$.add($r_i$), $accum \gets accum + utok(r_i)$\;
    
        \If{$i = n \text{ and } p \neq \emptyset$}{
            $\mathcal P$.add($p$)\tcp*{Leftover batches}
            \For{$o = 1,...,OL(R_t)$}{
                $\mathcal D$.add($d$)\;
            }
        }
    }
}
\end{algorithm}

\begin{definition}[Priority Estimation Model, PEM]

For a {\relquery} $R_t$ composed of multiple requests, the Priority Estimation Model (PEM) estimates $R_t$'s execution duration through two steps. First, PEM simulates the inference process on $R_t$, breaking it down into a sequence of prefill batches $\mathcal P=\{p_1, p_2,..., p_n\}$ and decode batches $\mathcal D=\{d_1, d_2,...,d_n\}$ using the Batch Decomposition Algorithm. Second, PEM applies the linear predictors $L^{prefill}$/$L^{decode}$ to estimate the execution duration of each prefill/decode batch, respectively. Finally, the total estimated duration of $R_t$ is the cumulative duration of all batches. Formally, we have:

\begin{equation}
    \begin{aligned}
        &\text{PEM}(R_t)=\sum_{p\in \mathcal P} L^{prefill}(p) + \sum_{d\in \mathcal{D}} L^{decode}(d),\\
        &\mathcal P, \mathcal D= \text{batch\_decompose}(R_t).
    \end{aligned}
\end{equation}

\end{definition}

\noindent \textbf{Fast Priority Estimation} The Priority Estimation Model described above offers accurate priority values but suffers from high computational overhead. To enable efficient priority estimation, we adopt two optimizations: (1) replacing the costly $utok(\cdot)$ computation with a more lightweight approximation $utok^*(\cdot)$, and (2) reusing previously computed priorities when updates are unnecessary.

First, it is computationally expensive to count the number of uncached tokens of all requests given the prefix cache content. In a typical LLM service, the LLM engine manages tens of thousands of tokens' KV cache with either hash tables~\cite{vllm} or radix trees~\cite{sglang}. Unfortunately, when queried against up to thousands of requests, the overhead of uncached tokens counting actually slows down the end-to-end LLM inference process according to our experiments. 

Therefore, we obtain the number of uncached tokens of a request $r$ with the approximation $utok^*(r)$, which is used to replace $utok(r)$. Concretely, we estimate the number of uncached tokens with the prefix cache miss ratio, which is calculated by sampling a small number of randomly sampled requests in the {\relquery} $R_t$. Since requests in the same {\relquery} share the same prompt template and data source, the prefix cache miss ratio calculated from the sampled requests is close to that of the whole. We denote the set of randomly sampled requests as $R_t^s$ whose size is a hyperparameter and the calculation of $utok^*(r)$ is defined in Equation~\ref{eq:cache_miss}. We calculate $\text{cache\_miss\_ratio}^{R_t}$ for each {\relquery} $R_t$ at the beginning of iteration $t$ and then reuse it for later uncached tokens counting.
Formally, we have:
\begin{equation}
\label{eq:cache_miss}
\begin{aligned}
    &\text{cache\_miss\_ratio}^{R_t} = \frac{\sum_{r\in R^s_t}utok(r)}{\sum_{r\in R_t^s} tok(r)}, R^s_t\subseteq R_t \\
    &utok^*(r) = tok(r) \times  \text{cache\_miss\_ratio}^{R_t} , r\in R_t
\end{aligned}
\end{equation}

Second, we reuse the previously computed priority value for a {\relquery} $R_t$ when the change of the runtime information hardly influences $R_t$'s execution duration. Concretely, if all requests of $R_t$ and $R_{t-1}$ are in the waiting queue $Q_t^-$, we can reuse the priority value of $R_{t-1}$ directly for $R_t$. That is,
\begin{equation}
    reuse(R_t, R_{t-1}) = \begin{cases}
    \text{True, if } \forall r \in R_{t-1}\cup R_t, r \in Q_t^-,  \\
    \text{False, otherwise. }
    \end{cases}
\end{equation}

This reuse is supported by the following two facts. (1) There is no progress difference between $R_t$ and $R_{t-1}$ since all requests in them are waiting. (2) Although the currently executing {\relquery} changes the prefix cache content, the currently executing {\relquery} is different from $R_t$ and they have different task templates and data sources. Therefore, the change in runtime information hardly influences the execution duration for the waiting $R_t$, and we can safely reuse the previous priority value.

Overall, DPU delivers fast and accurate priority estimation for {\relqueries} and effectively mitigates HoL blocking during both core and tail running times as will be demonstrated in Section~\ref{sec:exp:ablation}.

\noindent \textbf{Starvation Prevention.} Naively applying the priority-based scheduling may lead to starvation for {\relqueries} with large workloads, whose users may wait very long to receive responses. To avoid the starvation problem, we propose to use $\text{unit\_waiting\_time}$ as a fairness metric for {\relquery} services, considering that {\relqueries} have different numbers of requests. The $\text{unit\_waiting\_time}$ of a {\relquery} $R_t$ is calculated by dividing the waiting time of $R_t$ by the number of requests contained in $R_0$. That is,
\begin{equation}
    \text{unit\_waiting\_time}(R_t) = \text{waiting\_time}(R_t) / req(R_0).
\end{equation}

At each iteration $t$, after DPU performs priority updating, we additionally check the $\text{unit\_waiting\_time}$ of each waiting {\relquery}. If $\text{unit\_waiting\_time}(R_t)$ exceeds a predefined threshold, we will set $\texttt{Prio}(R_t)$ to zero to prioritize its execution and avoid starvation.

\subsection{Adaptive Batch Arranger}
\label{sec:method:apa}

The Adaptive Batch Arranger (ABA) controls the prefill and decode phase arrangement given the priority information. More concretely, at each iteration $t$, we have two candidate batches to execute, namely a candidate decode batch $d^{cand}_t$ that contains running requests and a candidate prefill batch $p^{cand}_t$ that contains waiting requests. The execution of $d^{cand}_t$ generates next tokens for running requests and may complete some of them. The execution of $p^{cand}_t$ starts the generation of the waiting requests contained, populates their KV cache, and outputs their first tokens.

Intuitively, the relative order of execution of $p^{cand}_t$ and $d^{cand}_t$ depends on the priority values of the requests contained. By checking and comparing the smallest priority value in each candidate batch, we can tell the current inference situation, which could be (1) the running {\relquery} is being preempted by a high-priority {\relquery} that just arrives, (2) the running {\relquery} is in the middle of execution without interruption, and (3) the running {\relquery} almost finishes and the later low-priority {\relqueries} are waiting in the queue. ABA has different arrangements in these situations.

Next, we present the construction of the two candidate batches and discuss in detail how to choose one candidate to execute in this iteration to minimize the average latency.

The candidate decode batch $d^{cand}_t$ contains all requests in the running queue $Q^+_t$. The reason is that the decoding process is memory-bound and it is beneficial to maximize the decoding batch size to accelerate the LLM inference, as noted in previous works~\cite{vllm, agrawal2024taming}.

The candidate prefill batch $p^{cand}_t$ is constructed by iteratively adding requests from the front of the waiting queue $Q_t^-$. Since requests are sorted by priority values in ascending order in the waiting queue, we can ensure that {\relqueries} with lower priority values are prefilled earlier to avoid HoL blocking. The iterative requests adding must comply with the three constraints as we have introduced in Algorithm~\ref{algo:batch_decompose}. Furthermore, we restrict that $p^{cand}_t$ includes requests from only one {\relquery}, which ensures that we can always detect the transition boundary for optimization when moving from one {\relquery} to the next.

\noindent \textbf{Empty Candidate Cases.} When both $d^{cand}_t$ and $p_t^{cand}$ are empty, the inference engine is idle and waiting for requests to arrive. When one of them is empty, ABA executes the non-empty candidate. 

In other cases where both candidates are not empty, ABA determines which candidate to execute in this iteration based on priority information to minimize average latency. We denote the smallest priority value of requests in $d^{cand}_t$ and $p^{cand}_t$ as $m^+_t$ and $m^-_t$, respectively. Formally, we have:
\begin{equation}
    \begin{aligned}
        m^+_t &= \min (\texttt{Prio}(r)), r\in d^{cand}_t, \\
        m^-_t &= \min (\texttt{Prio}(r)), r\in p^{cand}_t. \\
    \end{aligned}
\end{equation}

\noindent \textbf{{\relquery} Preemption.} When $m^+_t > m^-_t$, the first waiting {\relquery} in the waiting queue has a shorter execution duration than the running {\relqueries}. To avoid HoL blocking and minimize average latency, ABA pauses the decoding of running {\relqueries} and executes $p^{cand}_t$ in this iteration. Therefore, running but longer {\relqueries} are preempted by the waiting but shorter {\relquery}.

\noindent \textbf{Internal {\relquery} Execution.} When $m^+_t = m^-_t$, it means that we are in the middle of executing requests from the same {\relquery}. To minimize the core running time of this {\relquery}, we prioritize $p^{cand}_t$ over $d_t^{cand}$ to maximize the decoding batch size. 

\begin{figure}[t]
    \centering
    \includegraphics[width=0.8\linewidth]{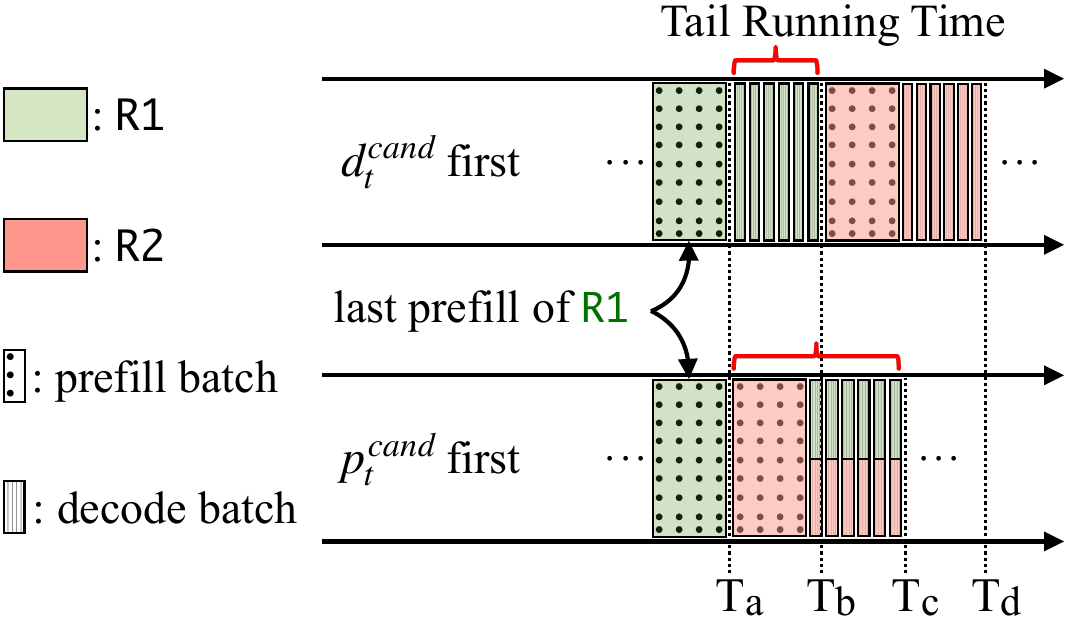}
    \vspace{-3ex}
    \caption{Comparison between prioritizing prefill candidate and prioritizing decode candidate in tail running time.}
    \label{fig:delta_latency}
\end{figure}

\noindent \textbf{Transitional {\relquery} Execution.} 
When $m^+_t < m^-_t$, the running {\relquery} has completed all its prefill batches but still has some unfinished decode batches. And here comes the latency trade-off between the almost finished running {\relquery} and later waiting {\relqueries}. We demonstrate the trade-off with an example in Figure~\ref{fig:delta_latency}. Without loss of generality, we consider two adjacent {\relqueries}, namely $R1$ that is finishing and $R2$ that is starting.

$T_a$ marks the end of the last prefill of $R1$. If we want to minimize the tail running time of $R1$, we should first execute $d^{cand}_t$ that contains requests of $R1$ as in the upper part of Figure~\ref{fig:delta_latency}. However, this arrangement misses the opportunity to combine the decoding of $R1$ and $R2$. By contrast, if we first execute $p^{cand}_t$ that contains requests of $R2$, we can combine the decoding of $R1$ and $R2$ and reduce the latency of $R2$. 
It is worth noting that combining $R1$ and $R2$ during decoding results in a larger decoding batch size, thereby improving GPU utilization, as we have discussed in Section~\ref{sec:bg:llm_infernce}. Consequently, prioritizing $p_t^{cand}$ enables more efficient GPU usage and leads to a shorter completion time for the two {\relqueries}, namely $T_c$ < $T_d$ as depicted in Figure~\ref{fig:delta_latency}.

Quantitatively, if we execute $p^{cand}_t$ earlier than $d_t^{cand}$, we \textbf{increase} the latency of $R1$ by $T_c-T_b$ and \textbf{decrease} the latency of $R2$ by $T_d-T_c$. We denote the change in the total latency of $R1$ and $R2$ as $\Delta_t = 2T_c-T_b-T_d$. If $\Delta_t < 0$, executing $p^{cand}_t$ first reduces average latency. Otherwise, executing $d^{cand}_t$ is better. The exact value of $\Delta_t$ depends on the involved batches in the current iteration.

We now formally compute the change in total latency $\Delta_t$ when executing the candidate prefill batch $p^{cand}_t$ ahead of the candidate decode batch $d_t^{cand}$ at iteration $t$. This scheduling impacts both the running {\relqueries} $\mathcal R^+_t$ and waiting {\relqueries} $\mathcal R^-_t$.
The overall latency change is given by $\Delta_t = \Delta^+_t + \Delta^-_t$, where $\Delta^+_t$ and $\Delta^-_t$ denote the changes for running and waiting {\relqueries}, respectively.

For running {\relqueries} $\mathcal R^+_t$, executing $p^{cand}_t$ introduces additional latency $\Delta^+_t$ because decoding is paused during the execution of $p^{cand}_t$. Additionally, incorporating more requests into future decode batches increases their execution time.
The total latency increase for running {\relqueries} is:
\begin{equation}
\label{eq:increased_latency}
    \begin{aligned}
     \Delta^+ &= L^{prefill}(p^{cand}_t)\times |\mathcal R^+_t| \\
     &+ \sum_{R\in \mathcal R^+_t} \alpha^d \times req(p^{cand}_t) \times \min(OL(R), OL(p^{cand}_t)).
    \end{aligned}
\end{equation}

For all waiting {\relqueries} $\mathcal R^-_t$, prioritizing $p^{cand}_t$ enables earlier combined decoding, reducing the latency of the first {\relquery} in $\mathcal{R}^-_t$. Earlier completion of this first {\relquery} also shortens the waiting time of subsequent requests. The total latency reduction $\Delta^-_t$ is computed as:
\begin{equation}
\begin{aligned}
     \Delta_t^- = |\mathcal R_t^-| \times \beta^d \times \min(OL(p^{cand}_t), 
     \max \{{OL(R), R\in\mathcal{R}^+_t}\}).
\end{aligned}
\end{equation}

Hence, we have the total latency change $\Delta_t$ as follows: 
\begin{equation}
\begin{aligned}
     \Delta_t = \Delta_t^+ + \Delta_t^-.
\end{aligned}
\end{equation}
If $\Delta_t < 0$, ABA executes $p^{cand}_t$ in this iteration. Otherwise, ABA executes $d_t^{cand}$ first. In summary, ABA adaptively determines whether to execute a prefill batch or a decode batch in the current iteration based on priority information and the {\relqueries} status in the inference engine, thus minimizing the average latency.

\section{Experiments}  
\label{sec:exp}

\subsection{Setups}
\label{sec:exp:setup}

\noindent \textbf{Implementation Details.} We implement {\system} on top of the state-of-the-art open-source LLM serving system vLLM~\cite{vllm} to benefit from existing optimizations in the LLM community, such as FlashAttention~\cite{dao2022flashattention}. 
We extend the request data type of vLLM by adding an attribute $\mathtt{rel\_id}$, which indicates the belonging {\relquery}, and an attribute $\mathtt{priority}$, which indicates the priority value, to facilitate {\relquery}-level priority estimation.
We replace the original FCFS scheduler of vLLM with a new scheduler where we insert our Dynamic Priority Updater and Adaptive Batch Arranger methods. At each iteration, the new scheduler additionally updates priority values, manages waiting and running queues, and decides the next batch as described in Figure~\ref{fig:arch}.
We use NCCL~\cite{NCCL} for tensor parallel communication when serving LLMs with more than one GPU.

\noindent \textbf{Hardware and Models.}
We conduct all the experiments on a server with four NVIDIA A100 (40GB) GPUs, where the NVLink connections are equipped for inter-GPU communication. Following existing works~\cite{vllm, agrawal2024taming,zhong2024distserve}, we choose OPT-13B~\cite{zhang2022opt}, Qwen2.5-32B~\cite{bai2023qwen}, and Llama2-70B~\cite{touvron2023llama} for experiments since they are representative models widely used in academia and industry. The OPT-13B model uses the traditional Multi-Head Attention (MHA) for the self-attention module and the other two models use Grouped-Query Attention (GQA) which further reduces KV cache usage. Following existing works~\cite{vllm,zhong2024distserve}, we use FP16 for all models. Detailed model sizes and hardware configurations are given in Table~\ref{tb:models}.

\noindent \textbf{Datasets and Workloads.}
Following existing work~\cite{liu2025optimizing}, we construct datasets by manually crafting five types of {\relqueries} over four real-world datasets collected from traditional machine learning tasks, yielding twenty different {\relquery} templates as follows. 

First, we collect four real-world datasets as shown in Table~\ref{tb:datasets}. The Amazon dataset~\cite{amazon_ds} contains product descriptions and user comments from the Amazon website. The Rotten dataset~\cite{rotten_ds} contains descriptions about movies and user reviews collected from the Rotten Tomatoes Website. The Beer dataset~\cite{beer_ds} contains comprehensive beer descriptions, ratings, and reviews from the RateBeer website. The PDMX dataset~\cite{pdmx_ds} collects the detailed metadata of public domain music. For each of the four datasets, we truncate and retain the first 10k rows for {\relquery} construction. 

Second, we prepare five types of queries corresponding to common tasks, including data filtering, data classification, numerical regression, text summarization, and open-ended question-answering. These five types support common semantic operators~\cite{patel2024semantic} individually or collectively as well as general LLM usage. Every query type will be adapted to fit the dataset context, creating the {\relquery} template. The output length limits for these five types are 5, 10, 5, 50, and 100 tokens, respectively. We present exemplar {\relquery} templates for these types in Table~\ref{tb:queries}.

Following existing works~\cite{vllm, agrawal2024taming, zhong2024distserve}, we construct serving traces by sampling {\relqueries} randomly. Concretely, we sample 100 {\relqueries} in each setting and the number of requests of each {\relquery} is randomly sampled within the (1,100) range, representing varied sizes of {\relqueries} sent by users in the real world. In total, each serving trace contains approximately 5k LLM requests. Following existing work~\cite{agrawal2024taming, zhong2024distserve}, we generate the {\relqueries} arrivals using the Poisson distribution with different request rates.  

\noindent \textbf{Baselines.} We compare {\system} to the following three baselines. 
\vspace{-3ex}
\begin{itemize}
    \item \textbf{vLLM}. vLLM~\cite{vllm} is a widely adopted LLM inference framework that uses the FCFS scheduling and prefill-prioritization arrangement. vLLM incorporates sophisticated optimization techniques such as PagedAttention, iteration-level batching, and prefix caching to accelerate inference speed. 
    \item \textbf{Sarathi}. Sarathi~\cite{agrawal2024taming} is a recent framework based on vLLM, which also uses FCFS scheduling but features the chunked prefill arrangement. Chunked prefill improves GPU utilization by breaking large prefill batches into smaller pieces and mixing the pieces with decode batches. 
    \item \textbf{vLLM-SP}. vLLM-SP is the state-of-the-art {\relquery} serving system we constructed that incorporates the \underline{S}tatic \underline{P}riority scheduling~\cite{zheng2023response,aiops2024qiu,zhao2024alise,fu2024efficient} into vLLM, which avoids the {\ProblemAbbr} in the waiting time. vLLM-SP shares the same code base as {\system} while disabling dynamic priority update and adaptive batch arrangement. 
\end{itemize}

\begin{table}[t]
    \caption{Models and hardware for experiments.}
    \centering
    \vspace{-3ex}
    \begin{tabular}{cccccc}
    \toprule
    Models & Abbr. & Sizes & Attn. & GPUs & Total Mem.\\
    \midrule
    \midrule
    OPT-13B & OPT & 24GB & MHA & A100 & 40GB\\
    Qwen2.5-32B & Qwen & 61GB & GQA & 2$\times$A100 & 80GB \\
    Llama2-70B & Llama & 129GB & GQA & 4$\times$A100 & 160GB \\
    \bottomrule
    \end{tabular}
    \label{tb:models}
\end{table}

\begin{table}[t]
    \caption{Datasets for experiments.}
    \vspace{-3ex}
    \centering
    \begin{tabular}{ccc}
    \toprule
    Datasets & Avg. Input Length & Avg. Output Length \\
    \midrule
    \midrule
    Amazon & 234 & 18 \\
    Rotten & 215 & 21 \\
    Beer & 174 & 19 \\
    PDMX & 158 & 23 \\
    \bottomrule
    \end{tabular}
    \label{tb:datasets}
\end{table}

Note that the key designs of {\system}, namely the Dynamic Priority Updater and the Adaptive Batch Arranger, are generally applicable to different LLM inference engines. In this paper, we make baselines and {\system} share the same vLLM code base to exclude the influence of implementations for a fair comparison.

\begin{table}[t]
    \caption{Exemplar {\relquery} templates for experiments.}
    \vspace{-3ex}
    \centering
    \begin{tabular}{r|l}
    \toprule
    Type & Example\tablefootnote{The
    \texttt{\{\}} in the example represents an attribute name.} \\
    \midrule
    \midrule
    Filter- &  \verb|Decide whether this movie is suitable|  \\ 
    ing &  \verb|for children based on the synopsis {}...| \\
    \midrule
    Classif- & \verb|Categorize the sentiment of the review {}| \\
    ication & \verb|as Negative, Positive, or Neutral...|\\
    \midrule
    Rating & \verb|Predict the user's rating on the beer| \\
    & \verb|based on producer {} and comment {}...|\\
    \midrule
    Summar- & \verb|Summarize the user's review {} on the| \\
    ization & \verb|product {} with in 20 words and ...|\\
    \midrule
    Open & \verb|Who are the most likely audiences| \\
    &\verb|for the music given its description {}...|\\
    \bottomrule
    \end{tabular}
    \label{tb:queries}
\end{table}

\begin{figure*}[t]
    \centering
    \includegraphics[width=\linewidth]{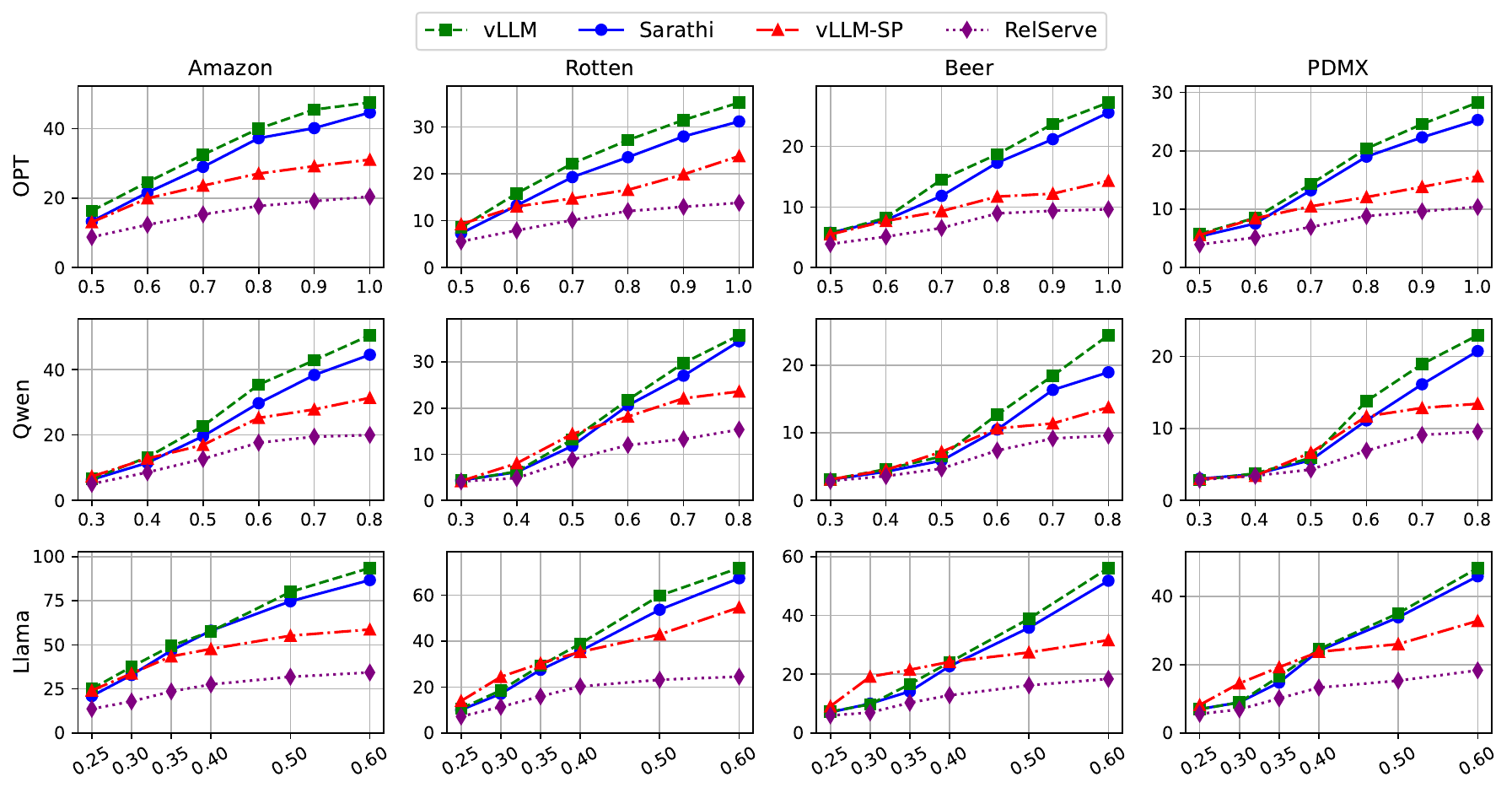}
    \vspace{-6ex}
    \caption{Latency comparison between baselines and {\system}. X-axis: Workload (unit: {\relqueries}/s). Y-axis: Latency (unit: s).}
    \label{fig:main_exp}
    \vspace{-3ex}
\end{figure*}

Recently, the disaggregated inference technique~\cite{zhong2024distserve,dynamo} was introduced to reduce token-level latency for chatbot services. However, this technique suffers from low hardware utilization~\cite{liang2025injecting} and is unsuitable for {\relquery} services. Under a common setting\footnote{Four A100 (40GB). Amazon dataset. 1.0 {\relquery}/s. Llama2-13B model.}, vLLM (agg-4TP)\footnote{The default vLLM (agg-4TP) uses aggregated inference with four-GPU tensor parallel. vLLM (dis-xPyD) variants adopt disaggregated inference using x prefill GPUs and y decode GPUs with NVIDIA Dynamo~\cite{dynamo}. Reconfiguring x and y at runtime does not help because switching a GPU between prefill and decode roles takes over 200s, comparable to or exceeding the end-to-end service duration in our experiments.}, vLLM (dis-1P3D), vLLM (dis-2P2D), and vLLM (dis-3P1D) achieve average {\relquery} latencies of 18.64s, 35.91s, 55.57s, and 141.01s, respectively. Consequently, we use aggregated inference for all frameworks in experiments for better performance.

\noindent \textbf{Metrics.} We focus on the average latency of {\relquery} services. In each setting, we measure the latency of {\relqueries} and calculate the average latency after the service completes. We run experiments in each setting three times and report the mean value.

\subsection{Main Results}
\label{sec:exp:main}

Figure~\ref{fig:main_exp} presents the overall latency results. {\system} consistently achieves the lowest average latency among all frameworks and settings, which verifies the effectiveness of our method. For the largest workload tested in each setting, {\system} reduces latency by 2.3 to 3.1 times compared to vLLM and 1.4 to 2.2 times compared to vLLM-SP. We discuss the details as follows.

First, in all settings, the improvement of {\system} over baselines increases with larger workloads. When the {\relquery} arrival rate is low, the execution timelines of different {\relqueries} are less overlapped, and there are fewer {\relqueries} in each iteration. Therefore, changing the execution order of {\relqueries} has less influence and the latencies of the three frameworks are similar. However, when the {\relquery} arrival rate is high, there are more {\relqueries} with different sizes in each iteration and the advantages of priority scheduling are revealed.

Second, compared to vLLM and Sarathi, both of which use FCFS scheduling and have similar performances, vLLM-SP uses static priority scheduling to avoid {\ProblemAbbr} in the waiting time. The static priority scheduling reduces average latency by 1.6 times across all settings for the largest workload. However, vLLM-SP cannot address {\ProblemAbbr} in the core and tail running times. {\system} takes the opportunity to tackle the {\ProblemAbbr} problem comprehensively and further reduces the latency by 1.6 times compared to vLLM-SP. 

Third, with the largest model Llama2-70B, {\system} shows the most significant improvement over baselines, namely 2.8 times average latency reduction compared to vLLM and 1.9 times average latency reduction compared to vLLM-SP across the four datasets. Since larger LLMs have more parameters, they require more time to process the same requests and prefill/decode batches. As a result, inappropriate request ordering and prefill-decode batch arrangement lead to more severe HoL blocking, amplifying the performance gap between {\system} and the baselines.

Fourth, on Amazon, Rotten, Beer, and PDMX, {\system} achieves average latency reductions of 2.5, 2.6, 2.8, and 2.6 times, respectively when compared to vLLM, and reduces average latency by 1.6, 1.8, 1.5, and 1.6 times compared to vLLM-SP across all three models. Although these datasets have different input and output lengths, which leads to different prefill-decode batch arrangement requirements, {\system} can arrange the batch execution adaptively and minimize latency in each scenario.

\subsection{Ablation Studies}
\label{sec:exp:ablation}

\begin{figure}[t]
    \centering
    \includegraphics[width=\linewidth]{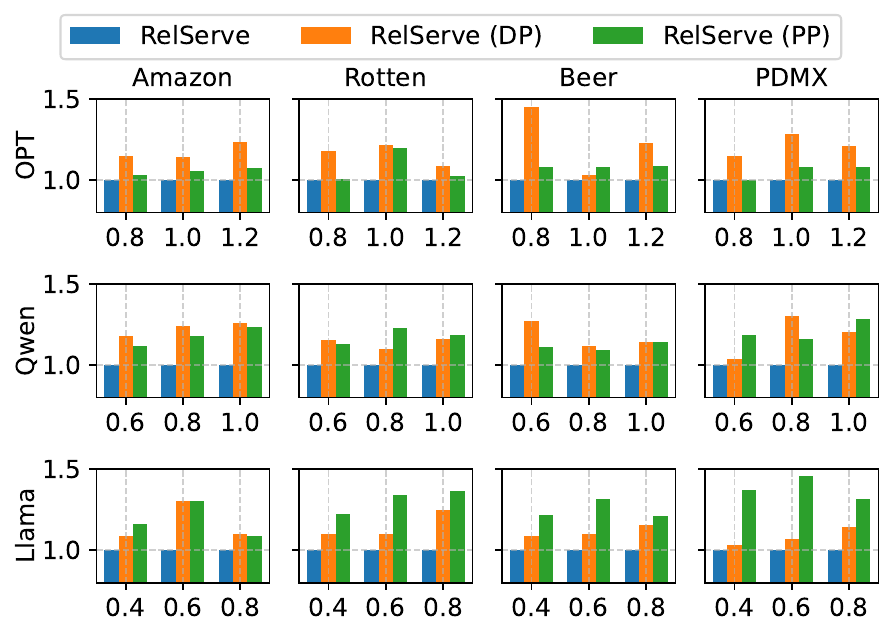}
    \vspace{-6ex}
    \caption{Comparison of prefill and decode arrangements. X-axis: Workload in {\relqueries}/s. Y-axis: Normalized latency.}
    \label{fig:ablation:apa}
\end{figure}

First, we conduct additional experiments to verify the effectiveness of the Adaptive Batch Arranger in Figure~\ref{fig:ablation:apa}. Specifically, we prepare two variants of {\system}, namely {\system} (PP) and {\system} (DP). {\system} (PP) means that we always prioritize the candidate prefill batch of waiting {\relqueries} in the tail running time, which is the default behavior of vLLM. {\system} (DP) means that we always prioritize the candidate decode batch of the current running {\relqueries} in the tail running time, which is a na\"ive fix to the long tail running time problem. 

Overall, the {\system} achieves the lowest latency in all settings, which confirms the superiority of ABA. In addition, we find that the relative performance of {\system} (PP) and {\system} (DP) varies in different settings. For example, in the PDMX + Llama setting, the large model leads to a long execution time of the prefill batch, and the long output length of PDMX leads to more decode batches for each {\relquery}. Consequently, during tail running time, {\system} (PP) is faced with more severe interruption of prefill batches of later {\relqueries} when decoding running {\relqueries}, leading to its worse performance compared to {\system} (DP). But the situation is opposite for the OPT + Amazon setting, where the output length is short and the model is small. Considering the complex dynamics of datasets, models, and workloads, static prefill-decode arrangements cannot optimize the latency trade-off, and ABA is the rescue.

\begin{figure}[t]
    \centering
    \includegraphics[width=\linewidth]{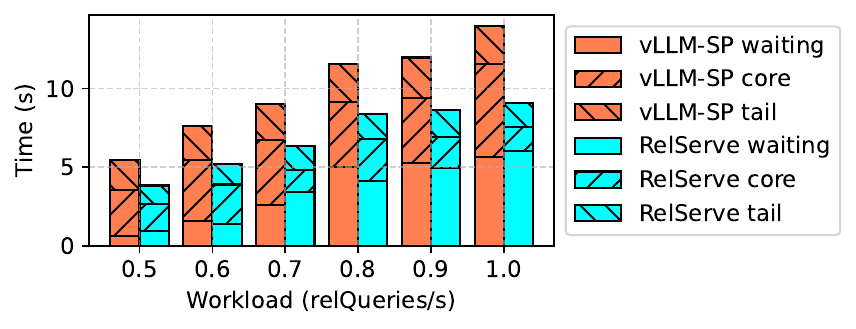}
    \vspace{-6ex}
    \caption{Breakdown of serving latency.}
    \label{fig:breakdown}
\end{figure}

Second, we break down the latency of vLLM-SP and {\system} into the three latency periods as shown in Figure~\ref{fig:breakdown}. The results are obtained with the OPT model and the Beer dataset and are similar in other settings. {\system} has a similar waiting time but considerably shorter core and tail running times compared to vLLM-SP. The reduced core running time reveals the effectiveness of DPU, which avoids HoL blocking with dynamic priority updating. The tail running time also decreases thanks to the APA arrangement.

Third, we study the overhead of the DPU and the ABA component in Table~\ref{tb:overhead}. The results are obtained with the OPT model and the Beer dataset and are similar in other settings. We find that the cumulative durations of ABA and DPU are less than 1\% of end-to-end {\relquery} serving latency. Such low overheads confirm the efficiency and practicality of {\system}.

\begin{figure}[t]
    \centering
    \includegraphics[width=0.9\linewidth]{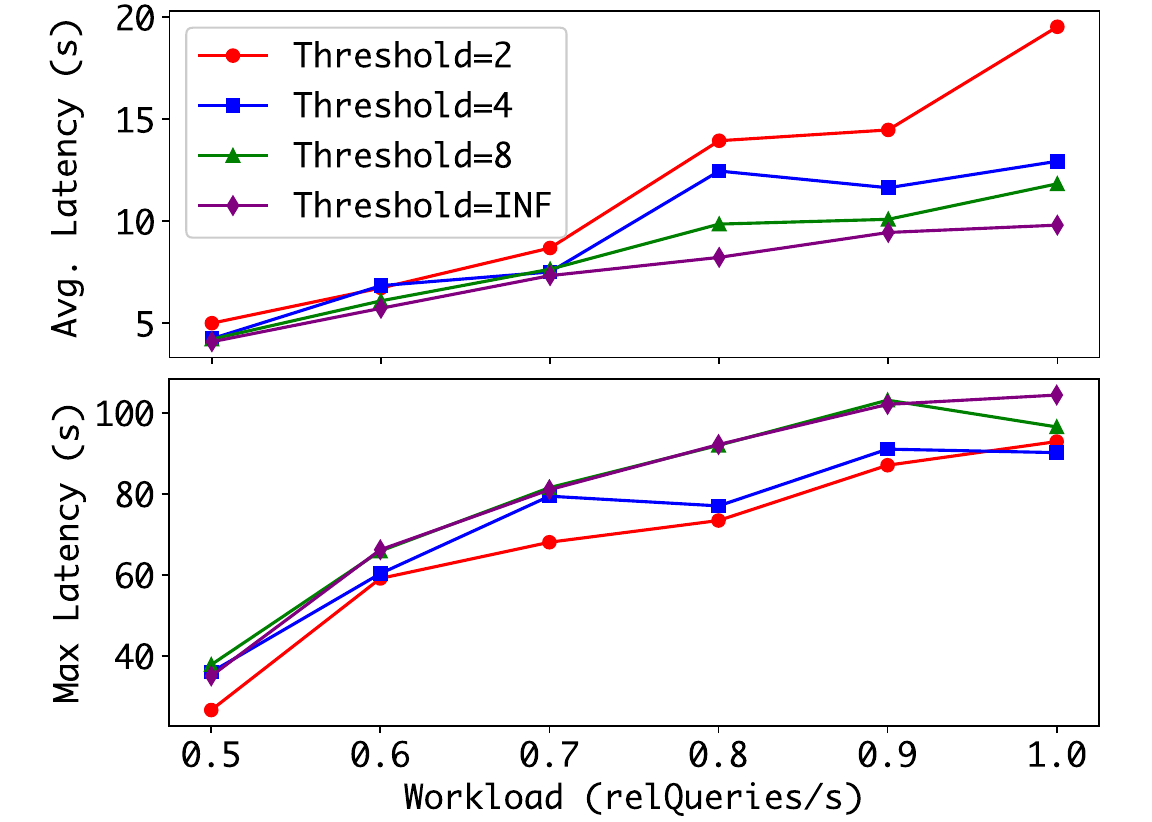}
    \vspace{-3ex}
    \caption{Influence of starvation thresholds (unit: s).}
    \label{fig:starvation}
\end{figure}

Fourth, we study the influence of starvation thresholds on the maximum and the average latency of {\system} as shown in Figure~\ref{fig:starvation}. We run this experiment with OPT model and Beer dataset and are similar in other settings. A lower threshold reduces the maximum latency by preventing {\relqueries} with large workloads from waiting too long, at the cost of a longer average latency. This trade-off should be optimized according to the preferences of users or applications. In addition, we observe that when the workload is large (e.g., 1.0) and the threshold is tight (e.g., 2), the server is congested and most of {\relqueries} have zero priority values. In this case, users should use more machines to serve the workload.

\begin{table}[t]
\caption{Overhead of DPU and ABA execution. E2E means end-to-end service duration. Unit of DPU/ABA/E2E: second.}
\vspace{-2ex}
\begin{tabular}{c|cccccc}
\toprule
Workload & 0.5    & 0.6    & 0.7    & 0.8    & 0.9    & 1.0    \\ 
\midrule
\midrule
DPU   & 0.45  & 0.33  & 0.36  & 0.51  & 0.41  & 0.40  \\ 
\midrule
ABA    & 0.19  & 0.14  & 0.21  & 0.22  & 0.12  & 0.12  \\ 
\midrule
E2E   & 202.2 & 171.1 & 158.1 & 156.4 & 152.6 & 146.7 \\ 
\bottomrule
\end{tabular}
\label{tb:overhead}
\end{table}


\section{Related Work}
\label{sec:related_work}
\noindent \textbf{LLM Inference Optimizations}. Inference optimizations~\cite{db1, db2, db3, db4, db5, db6, db7, db8, aptserve} are important when deploying machine learning models to real-world applications. The high cost of LLM attracts much attention to improve the inference performance. Orca~\cite{orca} introduces iteration-level batching that improves the throughput. vLLM~\cite{vllm} proposes paged-attention for efficient KV cache management. SARATHI~\cite{agrawal2024taming} introduces the chunked prefill method to improve the hardware utilization. However, these general LLM engines use FCFS scheduling that leads to severe HoL blocking.

\noindent \textbf{LLM Request Scheduling}. Some recent methods~\cite{zheng2023response, aiops2024qiu, zhao2024alise, fu2024efficient} propose to use priority scheduling to avoid the HoL blocking problem in LLM inference. They estimate the duration of one request by predicting exact or estimated output lengths and scheduling requests accordingly. These works use static priority, which works well in general chatbot workloads but causes inaccurate priority estimation and HoL blocking in {\relquery} workloads. 

\noindent \textbf{Relational LLM Inference}. LOTUS~\cite{patel2024semantic} accelerates semantic operators by selectively dispatching workload to proxy models, which are smaller and less accurate machine learning models that run faster than LLM. LOTUS applies only to semantic operators but not general {\relqueries}. OPHR~\cite{liu2025optimizing} accelerates offline LLM inference on relational data by re-constructing requests to maximize the prefix cache hit ratio. However, OPHR needs to modify the content of requests, which harms accuracy and requires prior knowledge about the database. Both LOTUS and OPHR are optimized for throughput in the offline processing setting. Unlikely, {\system} focuses on minimizing {\relquery} latency in the online serving setting without influencing accuracy. {\system} neither modifies the user-specified LLM model nor alters the original {\relquery} content, ensuring practicality compared to LOTUS and OPHR.

\vspace{-2ex}

\section{Conclusion}
\label{sec:conclusion}
This paper presents {\system}, an efficient framework for {\relquery} inference serving. We identify that the Head-of-Line blocking problem causes the prolonged {\relquery} serving latency and persists in the core and tail running times due to different reasons. First, the absence of a dynamic priority update mechanism causes HoL blocking during the core running period. Second, the na\"ive batch arrangement leads to HoL blocking in tail execution and fails to balance the latency trade-offs between running and waiting {\relqueries}. {\system} addresses these two problems by introducing the Dynamic Priority Updater that is both efficient and accurate, and the Adaptive Batch Arranger, which quantitatively evaluates and chooses from prefill and decode candidates for minimal latency. We extensively compare {\system} with existing inference engines using three mainstream LLM models, four real-world datasets, and five types of {\relqueries}. Experimental results show that {\system} reduces the average latency by up to 3.1 and 1.6 times compared to vLLM and state-of-the-art static priority scheduling methods.

\bibliographystyle{ACM-Reference-Format}
\bibliography{ref}

@article{achiam2023gpt,
  title={Gpt-4 technical report},
  author={Achiam, Josh and Adler, Steven and Agarwal, Sandhini and Ahmad, Lama and Akkaya, Ilge and Aleman, Florencia Leoni and Almeida, Diogo and Altenschmidt, Janko and Altman, Sam and Anadkat, Shyamal and others},
  journal={arXiv preprint arXiv:2303.08774},
  year={2023}
}

@article{touvron2023llama,
  title={Llama: Open and efficient foundation language models},
  author={Touvron, Hugo and Lavril, Thibaut and Izacard, Gautier and Martinet, Xavier and Lachaux, Marie-Anne and Lacroix, Timoth{\'e}e and Rozi{\`e}re, Baptiste and Goyal, Naman and Hambro, Eric and Azhar, Faisal and others},
  journal={arXiv preprint arXiv:2302.13971},
  year={2023}
}

@article{zhang2022opt,
  title={Opt: Open pre-trained transformer language models},
  author={Zhang, Susan and Roller, Stephen and Goyal, Naman and Artetxe, Mikel and Chen, Moya and Chen, Shuohui and Dewan, Christopher and Diab, Mona and Li, Xian and Lin, Xi Victoria and others},
  journal={arXiv preprint arXiv:2205.01068},
  year={2022}
}

@article{bai2023qwen,
  title={Qwen technical report},
  author={Bai, Jinze and Bai, Shuai and Chu, Yunfei and Cui, Zeyu and Dang, Kai and Deng, Xiaodong and Fan, Yang and Ge, Wenbin and Han, Yu and Huang, Fei and others},
  journal={arXiv preprint arXiv:2309.16609},
  year={2023}
}

@article{databricks,
  title={AI Functions on Databricks.},
  author={Databricks},
  url={https://docs.databricks.com/en/large-language-models/ai-functions.html},
  year={2025}
}

@article{bigquery,
  title={LLM with Vertex AI only using SQL queries in BigQuery.},
  author={Google Cloud Platform},
  url={https://cloud.google.com/blog/products/ai-machine-learning/llm-with-vertex-ai-only-using-sql-queries-in-bigquery},
  year={2024}
}

@article{redshift,
  title={Large Language Models for sentiment analysis with Amazon Redshift ML.},
  author={Amazon Web Services},
  url={https://aws.amazon.com/blogs/big-data/large-language-models-for-sentiment-analysis-with-amazon-redshift-ml-preview},
  year={2024}
}

@article{streaming_analytics,
  title={Leveraging streaming analytics for actionable insights with gen AI and Dataflow.},
  author={Google Cloud Platform},
  url={https://cloud.google.com/blog/products/data-analytics/learn-how-to-get-real-time-llm-insights-using-dataflow},
  year={2024}
}

@article{gptx_functions,
  title={GPTX Functions for Sheets.},
  author={Coefficient},
  url={https://help.coefficient.io/hc/en-us/articles/20602443535643-GPTX-Functions-for-Sheets},
  year={2025}
}

@article{sheetgpt,
  title={},
  author={Sheetgpt},
  url={https://www.sheetgpt.ai/},
  year={2025}
}

@article{coefficient,
  title={},
  author={Coefficient},
  url={https://www.Coefficient.io/},
  year={2025}
}

@article{cellm,
  title={},
  author={cellm},
  url={https://www.getcellm.com/},
  year={2025}
}

@article{TableauPulse,
  title={Get AI-Powered, Multi-Metric Exploration in Tableau Cloud with Tableau Pulse},
  author={Tableau},
  url={https://www.tableau.com/blog/tableau-pulse-enhanced-qa},
  year={2025}
}

@article{duckdb,
  title={Benchmarking DuckDB over Time.},
  author={DuckDB},
  url={https://duckdb.org/2024/06/26/benchmarks-over-time.html},
  year={2024}
}

@article{excel,
  title={Excel performance: Performance and limit improvements.},
  author={Microsoft},
  url={https://learn.microsoft.com/en-us/office/vba/excel/concepts/excel-performance/excel-performance-and-limit-improvements},
  year={2024}
}

@article{fastertransformer,
  title={Faster Transformer.},
  author={NVIDIA},
  journal={https://github.com/NVIDIA/FasterTransformer},
  year={2024}
}

@inproceedings{liu2025optimizing,
  title={Optimizing LLM Queries in Relational Data Analytics Workloads},
  author={Liu, Shu and Biswal, Asim and Kamsetty, Amog and Cheng, Audrey and Schroeder, Luis Gaspar and Patel, Liana and Cao, Shiyi and Mo, Xiangxi and Stoica, Ion and Gonzalez, Joseph E and others},
  booktitle={Eighth Conference on Machine Learning and Systems},
  year={2025}
}

@article{patel2024semantic,
  title={Semantic operators: A declarative model for rich, ai-based analytics over text data},
  author={Patel, Liana and Jha, Siddharth and Asawa, Parth and Pan, Melissa and Guestrin, Carlos and Zaharia, Matei},
  journal={arXiv preprint arXiv:2407.11418},
  pages={38},
  year={2024}
}

@inproceedings{vllm,
  title={Efficient memory management for large language model serving with pagedattention},
  author={Kwon, Woosuk and Li, Zhuohan and Zhuang, Siyuan and Sheng, Ying and Zheng, Lianmin and Yu, Cody Hao and Gonzalez, Joseph and Zhang, Hao and Stoica, Ion},
  booktitle={Proceedings of the 29th Symposium on Operating Systems Principles},
  pages={611--626},
  year={2023}
}

@article{aptserve,
author = {Gao, Shihong and Zhang, Xin and Shen, Yanyan and Chen, Lei},
title = {Apt-Serve: Adaptive Request Scheduling on Hybrid Cache for Scalable LLM Inference Serving},
year = {2025},
issue_date = {June 2025},
publisher = {Association for Computing Machinery},
address = {New York, NY, USA},
volume = {3},
number = {3},
url = {https://doi.org/10.1145/3725394},
doi = {10.1145/3725394},
journal = {Proc. ACM Manag. Data},
month = jun,
articleno = {130},
numpages = {28}
}

@inproceedings{amazon_ds,
  title={Ups and downs: Modeling the visual evolution of fashion trends with one-class collaborative filtering},
  author={He, Ruining and McAuley, Julian},
  booktitle={proceedings of the 25th international conference on world wide web},
  pages={507--517},
  year={2016}
}

@article{rotten_ds,
  title={Seeing stars: Exploiting class relationships for sentiment categorization with respect to rating scales},
  author={Pang, Bo and Lee, Lillian},
  journal={arXiv preprint cs/0506075},
  year={2005}
}

@inproceedings{beer_ds,
  title={Learning attitudes and attributes from multi-aspect reviews},
  author={McAuley, Julian and Leskovec, Jure and Jurafsky, Dan},
  booktitle={2012 IEEE 12th International Conference on Data Mining},
  pages={1020--1025},
  year={2012},
  organization={IEEE}
}

@inproceedings{pdmx_ds,
  title={PDMX: A Large-Scale Public Domain MusicXML Dataset for Symbolic Music Processing},
  author={Long, Phillip and Novack, Zachary and Berg-Kirkpatrick, Taylor and McAuley, Julian},
  booktitle={ICASSP 2025-2025 IEEE International Conference on Acoustics, Speech and Signal Processing (ICASSP)},
  pages={1--5},
  year={2025},
  organization={IEEE}
}

@inproceedings{aiops2024qiu,
  author  = {Qiu, Haoran and Mao, Weichao and Patke, Archit and Cui, Shengkun and Jha, Saurabh and Wang, Chen and Franke, Hubertus and Kalbarczyk, Zbigniew T. and Ba\c{s}ar, Tamer and Iyer, Ravishankar K.},
  title   = {Efficient Interactive LLM Serving with Proxy Model-based Sequence Length Prediction},
  year    = {2024},
  pages = {1--7},
  publisher = {Association for Computing Machinery},
  volume = {5},
  address = {San Diego, CA, USA},
  booktitle = {The 5th International Workshop on Cloud Intelligence / AIOps at ASPLOS 2024},
}

@inproceedings{zhao2024alise,
  title={ALISE: Accelerating Large Language Model Serving with Speculative Scheduling},
  author={Zhao, Youpeng and Wang, Jun},
  booktitle={Proceedings of the 43rd IEEE/ACM International Conference on Computer-Aided Design},
  pages={1--9},
  year={2024}
}

@article{fu2024efficient,
  title={Efficient LLM Scheduling by Learning to Rank},
  author={Fu, Yichao and Zhu, Siqi and Su, Runlong and Qiao, Aurick and Stoica, Ion and Zhang, Hao},
  journal={NeurIPS},
  year={2024}
}

@article{zheng2023response,
  title={Response length perception and sequence scheduling: An llm-empowered llm inference pipeline},
  author={Zheng, Zangwei and Ren, Xiaozhe and Xue, Fuzhao and Luo, Yang and Jiang, Xin and You, Yang},
  journal={Advances in Neural Information Processing Systems},
  volume={36},
  pages={65517--65530},
  year={2023}
}

@article{blazewicz1996job,
  title={The job shop scheduling problem: Conventional and new solution techniques},
  author={B{\l}a{\.z}ewicz, Jacek and Domschke, Wolfgang and Pesch, Erwin},
  journal={European journal of operational research},
  volume={93},
  number={1},
  pages={1--33},
  year={1996},
  publisher={Elsevier}
}

@inproceedings{agrawal2024taming,
  title={Taming $\{$Throughput-Latency$\}$ tradeoff in $\{$LLM$\}$ inference with $\{$Sarathi-Serve$\}$},
  author={Agrawal, Amey and Kedia, Nitin and Panwar, Ashish and Mohan, Jayashree and Kwatra, Nipun and Gulavani, Bhargav and Tumanov, Alexey and Ramjee, Ramachandran},
  booktitle={18th USENIX Symposium on Operating Systems Design and Implementation (OSDI 24)},
  pages={117--134},
  year={2024}
}

@article{vaswani2017attention,
  title={Attention is all you need},
  author={Vaswani, Ashish and Shazeer, Noam and Parmar, Niki and Uszkoreit, Jakob and Jones, Llion and Gomez, Aidan N and Kaiser, {\L}ukasz and Polosukhin, Illia},
  journal={Advances in neural information processing systems},
  volume={30},
  year={2017}
}

@article{zhong2024distserve,
  title={DistServe: Disaggregating Prefill and Decoding for Goodput-optimized Large Language Model Serving},
  author={Yinmin Zhong and Shengyu Liu and Junda Chen and Jianbo Hu and Yibo Zhu and Xuanzhe Liu and Xin Jin and Hao Zhang},
  journal={OSDI},
  year={2024}
}

@article{orca,
  title={Orca: A Distributed Serving System for Transformer-Based Generative Models},
  author={Gyeong-In Yu and Joo Seong Jeong, Geon-Woo Kim and Soojeong Kim and Byung-Gon Chun},
  journal={OSDI},
  year={2022}
}

@inproceedings{sglang,
author = {Zheng, Lianmin and Yin, Liangsheng and Xie, Zhiqiang and Sun, Chuyue et, al.},
title = {SGLang: efficient execution of structured language model programs},
year = {2025},
isbn = {9798331314385},
publisher = {Curran Associates Inc.},
address = {Red Hook, NY, USA},
booktitle = {Proceedings of the 38th International Conference on Neural Information Processing Systems},
articleno = {2000},
numpages = {27},
location = {Vancouver, BC, Canada},
series = {NIPS '24}
}

@article{dao2022flashattention,
  title={Flashattention: Fast and memory-efficient exact attention with io-awareness},
  author={Dao, Tri and Fu, Dan and Ermon, Stefano and Rudra, Atri and R{\'e}, Christopher},
  journal={Advances in neural information processing systems},
  volume={35},
  pages={16344--16359},
  year={2022}
}

@article{db1,
author = {Chang, Chaokun and Lo, Eric and Ye, Chunxiao},
title = {Biathlon: Harnessing Model Resilience for Accelerating ML Inference Pipelines},
year = {2024},
issue_date = {June 2024},
publisher = {VLDB Endowment},
volume = {17},
number = {10},
issn = {2150-8097},
url = {https://doi.org/10.14778/3675034.3675052},
doi = {10.14778/3675034.3675052},
journal = {Proc. VLDB Endow.},
month = jun,
pages = {2631–2640},
numpages = {10}
}

@article{db2,
author = {Yang, Zhihui and Wang, Zuozhi and Huang, Yicong and Lu, Yao and Li, Chen and Wang, X. Sean},
title = {Optimizing machine learning inference queries with correlative proxy models},
year = {2022},
issue_date = {June 2022},
publisher = {VLDB Endowment},
volume = {15},
number = {10},
issn = {2150-8097},
url = {https://doi.org/10.14778/3547305.3547310},
doi = {10.14778/3547305.3547310},
journal = {Proc. VLDB Endow.},
month = jun,
pages = {2032–2044},
numpages = {13}
}

@article{db3,
author = {Sirin, Utku and Idreos, Stratos},
title = {The Image Calculator: 10x Faster Image-AI Inference by Replacing JPEG with Self-designing Storage Format},
year = {2024},
issue_date = {February 2024},
publisher = {Association for Computing Machinery},
address = {New York, NY, USA},
volume = {2},
number = {1},
url = {https://doi.org/10.1145/3639307},
doi = {10.1145/3639307},
journal = {Proc. ACM Manag. Data},
month = mar,
articleno = {52},
numpages = {31}
}

@article{db4,
author = {Kang, Daniel and Mathur, Ankit and Veeramacheneni, Teja and Bailis, Peter and Zaharia, Matei},
title = {Jointly optimizing preprocessing and inference for DNN-based visual analytics},
year = {2020},
issue_date = {October 2020},
publisher = {VLDB Endowment},
volume = {14},
number = {2},
issn = {2150-8097},
url = {https://doi.org/10.14778/3425879.3425881},
doi = {10.14778/3425879.3425881},
journal = {Proc. VLDB Endow.},
month = oct,
pages = {87–100},
numpages = {14}
}

@inproceedings{db5,
  title={Accelerating scalable graph neural network inference with node-adaptive propagation},
  author={Gao, Xinyi and Zhang, Wentao and Yu, Junliang and Shao, Yingxia and Nguyen, Quoc Viet Hung and Cui, Bin and Yin, Hongzhi},
  booktitle={2024 IEEE 40th International Conference on Data Engineering (ICDE)},
  pages={3042--3055},
  year={2024},
  organization={IEEE}
}

@inproceedings{db6,
  title={Pp-stream: Toward high-performance privacy-preserving neural network inference via distributed stream processing},
  author={Liu, Qingxiu and Huang, Qun and Chen, Xiang and Wang, Sa and Wang, Wenhao and Han, Shujie and Lee, Patrick PC},
  booktitle={2024 IEEE 40th International Conference on Data Engineering (ICDE)},
  pages={1492--1505},
  year={2024},
  organization={IEEE}
}

@inproceedings{db7,
  title={InferTurbo: A scalable system for boosting full-graph inference of graph neural network over huge graphs},
  author={Zhang, Dalong and Song, Xianzheng and Hu, Zhiyang and Li, Yang and Tao, Miao and Hu, Binbin and Wang, Lin and Zhang, Zhiqiang and Zhou, Jun},
  booktitle={2023 IEEE 39th International Conference on Data Engineering (ICDE)},
  pages={3235--3247},
  year={2023},
  organization={IEEE}
}

@inproceedings{db8,
  title={Accelerating machine learning inference with probabilistic predicates},
  author={Lu, Yao and Chowdhery, Aakanksha and Kandula, Srikanth and Chaudhuri, Surajit},
  booktitle={Proceedings of the 2018 International Conference on Management of Data},
  pages={1493--1508},
  year={2018}
}

@inproceedings{arapakis2014impact,
  title={Impact of response latency on user behavior in web search},
  author={Arapakis, Ioannis and Bai, Xiao and Cambazoglu, B Barla},
  booktitle={Proceedings of the 37th international ACM SIGIR conference on Research \& development in information retrieval},
  pages={103--112},
  year={2014}
}

@article{li2015supporting,
author = {Li, Boduo and Diao, Yanlei and Shenoy, Prashant},
title = {Supporting scalable analytics with latency constraints},
year = {2015},
issue_date = {July 2015},
publisher = {VLDB Endowment},
volume = {8},
number = {11},
issn = {2150-8097},
url = {https://doi.org/10.14778/2809974.2809979},
doi = {10.14778/2809974.2809979},
journal = {Proc. VLDB Endow.},
month = jul,
pages = {1166–1177},
numpages = {12}
}

@inproceedings{zou2013flexquery,
  title={FlexQuery: An online query system for interactive remote visual data exploration at large scale},
  author={Zou, Hongbo and Schwan, Karsten and Slawinska, Magdalena and Wolf, Matt and Eisenhauer, Greg and Zheng, Fang and Dayal, Jai and Logan, Jeremy and Liu, Qing and Klasky, Scott and others},
  booktitle={2013 IEEE International Conference on Cluster Computing (CLUSTER)},
  pages={1--8},
  year={2013},
  organization={IEEE}
}

@article{agarwal2012blink,
  title={Blink and it's done: interactive queries on very large data},
  author={Agarwal, Sameer and Iyer, Anand P and Panda, Aurojit and Madden, Samuel and Mozafari, Barzan and Stoica, Ion},
  year={2012},
  publisher={Association for Computing Machinery (ACM)}
}

@article{chaudhuri2011overview,
  title={An overview of business intelligence technology},
  author={Chaudhuri, Surajit and Dayal, Umeshwar and Narasayya, Vivek},
  journal={Communications of the ACM},
  volume={54},
  number={8},
  pages={88--98},
  year={2011},
  publisher={ACM New York, NY, USA}
}

@article{liang2025injecting,
  title={Injecting Adrenaline into LLM Serving: Boosting Resource Utilization and Throughput via Attention Disaggregation},
  author={Liang, Yunkai and Chen, Zhangyu and Zuo, Pengfei and Zhou, Zhi and Chen, Xu and Yu, Zhou},
  journal={arXiv preprint arXiv:2503.20552},
  year={2025}
}

@article{dynamo,
  title={Dynamo Inference Framework},
  author={NVIDIA},
  url={https://docs.nvidia.com/dynamo/latest/},
  year={2025}
}

@article{NCCL,
  title={Nvidia Collective Communications Library (NCCL)},
  author={NVIDIA},
  url={https://developer.nvidia.com/nccl},
  year={2025}
}

\end{document}